\documentclass[12pt,preprint]{aastex}

\newcommand{\etal  }{{et al.} }
\newcommand{\msun}{\thinspace M_\odot}  
  
\newcommand{\vect}[1]{\mbox{\boldmath$#1$}}

\newcommand{\cm  }{\,{\rm cm}^{-3} } 
\newcommand{\jcm}{{\rm cm^2\,s^{-1}}}
\newcommand{\dfrac}[2]{{\displaystyle \frac{#1}{#2}}  }

\newcommand{\rh  }{r_{\rm H}} 
\newcommand{\rht }{\tilde{r}_{\rm H}} 
\newcommand{\rp  }{r_{\rm p}} 
\newcommand{\ro  }{a_{\rm p}} 
\newcommand{\me  }{\thinspace M_\oplus } 

\newcommand{\hs}{h}
\newcommand{\mj}{M_{\rm J}}
\newcommand{\rj}{r_{\rm Jup}}
\newcommand{\tl}[1]{\tilde{#1}}
\newcommand{\deft}{\tilde{t}_{\rm p}}

\shorttitle{Angular momentum accretion}
\shortauthors{Machida  \etal 2007}

\begin{document}
\title{Angular Momentum Accretion onto a Gas Giant Planet}

\author{Masahiro N. Machida\altaffilmark{1}, Eiichiro Kokubo\altaffilmark{2}, Shu-ichiro Inutsuka\altaffilmark{1}, and Tomoaki Matsumoto\altaffilmark{3}} 

\altaffiltext{1}{Department of Physics, Graduate School of Science, Kyoto University, Sakyo-ku, Kyoto 606-8502, Japan; machidam@scphys.kyoto-u.ac.jp, inutsuka@tap.scphys.kyoto-u.ac.jp}
\altaffiltext{2}{Division of Theoretical Astronomy, National Astronomical Observatory of Japan, Osawa, Mitaka,
Tokyo 181-8588, Japan; kokubo@th.nao.ac.jp}
\altaffiltext{2}{Faculty of Humanity and Environment, Hosei University, Fujimi, Chiyoda-ku, Tokyo 102-8160, Japan; matsu@i.hosei.ac.jp}

\begin{abstract}
We investigate the accretion of angular momentum onto a protoplanet
 system using three-dimensional hydrodynamical simulations. 
We consider a local region around a protoplanet in a protoplanetary disk
 with sufficient spatial resolution.
We describe the structure of the gas flow onto and around the protoplanet in
 detail.
We find that the gas flows onto the protoplanet system in the vertical
 direction crossing the shock front near the Hill radius of the
 protoplanet, which is qualitatively different from the picture
 established by two-dimensional simulations.
The specific angular momentum of the gas accreted by the protoplanet
 system increases with the protoplanet mass.
At Jovian orbit, when the protoplanet mass $M_{\rm p}$ is 
 $M_{\rm p}\lesssim 1\mj$, where $\mj$ is Jovian mass, the specific angular
 momentum increases as $j\propto M_{\rm p}$.  
On the other hand, it increases as $j\propto M_{\rm p}^{2/3}$ when the
 protoplanet mass is $M_{\rm p}\gtrsim 1\mj$.
The stronger dependence of the specific angular momentum on the
 protoplanet mass for $M_{\rm p}\lesssim 1\mj$ is due to thermal pressure of the gas.
The estimated total angular momentum of a system of a gas giant planet and a
 circumplanetary disk is two-orders of magnitude larger than those of
 the present gas giant planets in the solar system.
A large fraction of the total angular momentum contributes to the
 formation of the circumplanetary disk. 
We also discuss the satellite formation from the circumplanetary disk.
\end{abstract}

\keywords{accretion, accretion disks --- hydrodynamics --- planetary systems ---planets and satellites: formation--- solar system: formation}

\section{INTRODUCTION}
Until now, more than 200 extrasolar planets (or exoplanets) have been 
 detected mainly by measuring the radial motion of their parent star
 along the line of sight.  
Almost all exoplanets observed by this method are giant planets, like
 Jupiter and Saturn in our solar system, because massive planets are
 preferentially observed.  
Although these planets are supposed to be formed in the disk surrounding
 the central star (i.e., circumstellar disk or protoplanetary disk),
 their formation process has not been fully understood yet.  
In the core accretion scenario \citep{hayashi85}, a solid core or
 protoplanet with $\simeq 10\me$, where $\me$ is the Earth mass,
 captures a massive gas envelope from the protoplanetary disk by
 self-gravity to become a gas giant planet.

The evolution of the gaseous protoplanet has been studied with the approximation of  spherical symmetry  including radiative transfer
 \citep[e.g.,][]{mizuno80,bodenheimer86,pollack96,ikoma00}. 
\citet{ikoma00} showed that rapid gas accretion is triggered when the
 solid core mass exceeds $\simeq 5-20\me$, and the protoplanet quickly
 increases their mass by gas accretion. 
However, the angular momentum of the accreting gas was ignored in these
 studies, because they assumed spherical symmetry. 
Since the gas accretes onto the solid core with a certain amount of the 
 angular momentum, a circumplanetary disk forms around the protoplanet, analogously to the formation of a protoplanetary disk around a protostar.
The difference in the disk formation between the protostar and
 protoplanet is the region from which the central object acquires the
 angular momentum.  
The protostar acquires the angular momentum from a parent cloud, while
 the protoplanet acquires it from the shearing motion in the
 protoplanetary (circumstellar) disk.
In addition, the gravitational sphere of the protostar spreads almost 
 infinitely, while the gravitational sphere (i.e., the Hill sphere) of 
 the protoplanet is limited in the region around the protoplanet because 
 the gravity of the central star exceeds that of the protoplanet outside
 the Hill sphere.

The numerical simulations are useful to investigate the gas accretion
 onto a protoplanet and its circumplanetary disk (hereafter we just call
 them as a protoplanet system).
\citet{korycansky91} studied giant planet formation using one-dimensional
 quasi-spherical approximation with angular momentum transfer. 
They showed that as the protoplanet contracts, outer layers of the
 envelope containing sufficient specific angular momentum remain in 
 bound orbit, and form a circumplanetary disk. 
However, owing to the spherical symmetry, the accretion flow from the
 protoplanetary disk to the protoplanet could not be investigated in
 their study.  
To study the accretion flow onto a protoplanet and the acquisition
 process of the angular momentum in detail, a multidimensional
 simulation is necessary. 
\citet{sekiya87} investigated the gas flow around a protoplanet with relatively low resolution, and found
 that the spin rotation vector of the protoplanet becomes parallel to the orbital rotation vector. 
Recently, the flow pattern was carefully
 investigated by many authors 
 \citep{miyoshi99, lubow99, kley01, dangelo02, dangelo03}. 
However, since the main purpose of these studies was to clarify the
 planet migration process in a large scale (i.e., outside the Hill radius), they did not
 investigate the region near the protoplanet (i.e., inside the Hill
 radius) with sufficient resolution.  
Thus, in their simulations, we cannot study the gas stream inside the
 Hill radius.

To investigate the accretion flow onto the protoplanet system, we need
 to cover a large spatial scale from the region far from the Hill sphere
 to that in the proximity to the protoplanet. 
For Jupiter, since the Hill radius is $\rh = 744\,\rp$, where $\rp$ is
 Jovian radius, we have to resolve at least  $\sim$1000 times different
 scales.  
To cover a large dynamical range in scales, a few authors used the nested-grid
 method.
\citet{dangelo02,dangelo03} investigated the relation between the spiral
 patterns within the Hill radius and migration rate using
 three-dimensional nested-grid code.
Although they resolved the region inside the Hill radius, they did not
 investigate the structure in the proximity to the protoplanet because
 they adopted a sink cell at $0.1r_{\rm H}$ that corresponds to
 $\sim70\,\rp$ at Jovian orbit (5.2 AU).
Thus, we cannot observe a circumplanetary disk at $r\ll\rh$ in their calculation. 
\citet{tanigawa02a} also investigated the gas flow around a protoplanet using two-dimensional nested-grid code.  
They resolved the region from $12\,\rh$ to $0.005\,\rh$.
They found that a circumplanetary disk with $100\me$ is formed around
 the protoplanet.  
\citet{tanigawa02b} and \citet{machida06c} investigated the evolution of
 the protoplanet system using three-dimensional nested-grid code. 
They found that the gas flow pattern in three dimensions is qualitatively
 different from that in two dimensions: the gas is flowing into the
 protoplanet system only in the vertical direction in three-dimensional
 simulations.

In the present study, we calculated the evolution of the protoplanet system using three-dimensional nested-grid code. 
We found that after the flow around the protoplanet reaches a steady
 state, the angular momentum accreting onto the protoplanet system is 
 well converged regardless of both the cell width and the size of the
 sink cell region, while the mass accretion rate is not well converged. 
Although we calculated the evolution of the protoplanet system with spatial resolution much higher than previous studies, we still need further
 higher spatial resolution to determine the mass accretion rate onto the
 protoplanet.
Thus, in this paper, we focus on the gas flow onto and around the
 protoplanet system and the accretion process of the angular momentum, 
 and do not deal with the mass accretion rate (we plan to investigate the
 mass accretion rate with higher spatial resolution using a
 higher-performance computer in a subsequent paper).  
Note that the specific angular momentum accreting onto the protoplanet
 system with fixed protoplanet mass does not strongly depend on the mass accretion
 rate as described in the following sections.

The structure of the paper is as follows. 
The frameworks of our models are given in \S 2 and the numerical method
 is described in \S3. 
The numerical results are presented in \S 4.  
\S 5 is devoted for discussions.
We summarize our conclusions in \S6.

\section{MODEL}
\subsection{Master Equations}

We consider a local region around a protoplanet using shearing sheet
 model \citep[e.g.,][]{goldreich65}.  
We assume that the temperature is constant and the self-gravity of the
 disk is negligible. 
The orbit of the protoplanet is assumed to be circular on the equatorial
 plane of the circumstellar disk. 

We set up local rotating Cartesian coordinates with the origin at the
 protoplanet and the $x$-, $y$-, and $z$-axis are radial, azimuthal, and
 vertical direction of the disk, respectively.  
We solve the equations of hydrodynamics without self-gravity:
\begin{equation}
 \dfrac{\partial \rho}{\partial t} + \nabla \cdot (\rho \, \vect{v}) = 0,\\
 \label{eq:basic-1}
\end{equation}
\begin{equation}
 \dfrac{\partial \vect{v}}{\partial t} + (\vect{v} \cdot \nabla) \vect{v} =
  - \dfrac{1}{\rho} \nabla P - \nabla \Phi_{\rm eff} - 
  2 \vect{\Omega_{\rm p}} \times \vect{v},  
  \label{eq:basic-2}
\end{equation}
 where $\rho$, $\vect{v}$, $P$, $\Phi_{\rm eff}$, and
 $\vect{\Omega}_{\rm p}$ are the gas density, velocity, gas pressure,
 effective potential, and Keplerian angular velocity of the protoplanet,
 respectively.  
In the above equations, the curvature terms are neglected.
We adopt an isothermal equation of state, 
\begin{equation}
 P = c_{\rm s}^2 \rho,
\label{eq:basic-3} 
\end{equation}
 where $c_{\rm s}$ is the sound speed. 
The Keplerian angular velocity of the protoplanet is given by
\begin{equation}
\Omega_{\rm p} = \left( \dfrac{ G\, M_{\rm c} } { \ro^3} \right)^{1/2},
\label{eq:omegap}
\end{equation}
 where $G$, $M_{\rm c}$, and $\ro$ are the gravitational constant, mass
 of the central star, and orbital radius of the protoplanet, respectively. 
The effective potential $\Phi_{\rm eff}$  is given by
\begin{equation}
\Phi_{\rm eff} = - \dfrac{\Omega_{\rm p}^2}{2}(3 x^2 - z^2) \, - \,
 \dfrac{G M_{\rm p}}{r}, 
\label{eq:phi}
\end{equation}
 where $M_{\rm p}$ and $r$ are the mass of the protoplanet, and the
 distance from the center of the protoplanet \citep[e.g.,][]{miyoshi99}. 
The first term is composed of the gravitational potential of the central
 star and the centrifugal potential, and higher orders in $x$, $y$ and
 $z$ are neglected. 
The second term is the gravitational potential of the protoplanet.
Using the Hill radius 
\begin{equation}
 r_{\rm H} = \left( \dfrac{M_{\rm p}}{3M_{\rm c}} \right)^{1/3} \ro,
\label{eq:hill}
\end{equation} 
 equation~(\ref{eq:phi}) can be rewritten as 
\begin{equation}
\Phi_{\rm eff} = \Omega_{\rm p}^2 \left( - \dfrac{3 x^2 - z^2}{2} \, -
				   \, \dfrac{3\,r_{\rm H}^3}{r} \right). 
\label{eq:phi-eff}
\end{equation}

\subsection{Circumstellar Disk Model}

Our initial settings are similar to \citet{miyoshi99} and \cite{machida06c}. 
The gas flow has a constant shear in the $x$-direction as
\begin{equation}
\vect{v_0} = ( 0,\,  -\dfrac{3}{2}\Omega_{\rm p}\, x, \,0 ).
\label{eq:shear}
\end{equation}
For hydrostatic equilibrium, the density is given by 
\begin{equation}
\rho_0 = \dfrac{\sigma_0}{\sqrt{2\pi}h} {\rm exp } \left(- \dfrac{
						    z^2}{2 h^2} \right), 
\end{equation}
 where $\sigma_0$ ($\equiv \int_{-\infty}^{\infty} \rho \, dz $) is the
 surface density of the unperturbed disk. 
The scale height $h$ is related to the sound speed $c_{\rm s}$ as
 $h=c_{\rm s}/\Omega_{\rm p}$. 

In the standard solar nebular model \citep{hayashi81,hayashi85}, the
 temperature $T$, sound speed $c_{\rm s}$, and gas density $\rho_0$ are
 given by
\begin{equation}
T = 280 \left( \dfrac{L}{L_{\odot}} \right)^{1/4}
 \left(\dfrac{\ro}{1\,{\rm AU}} \right)^{-1/2}, 
\label{eq:nebular-temp}
\end{equation}
 where $L$ and $L_{\odot}$ are the protostellar and solar luminosities,  
\begin{equation}
 c_{\rm s} = \left( \dfrac{k\,T}{\mu m_{\rm H}} \right)^{1/2} =
 1.9\times 10^4\, \left( \dfrac{T}{10\,{\rm K}} \right)^{1/2} \, \left(
		    \dfrac{2.34}{\mu} \right)^{1/2} \ \  {\rm cm\,s^{-1}}, 
\label{eq:nebular-cs}
\end{equation}
 where $\mu =2.34$ is the mean molecular weight of the gas composed
 mainly of H$_2$ and He, and   
\begin{equation}
\rho_0 = 1.4 \times 10^{-9} \left( \dfrac{\ro}{1\,{\rm AU}}
			    \right)^{-11/4} \ \ {\rm g}\cm, 
\label{eq:nebular-dens}
\end{equation}
respectively.
When $M_c = 1\msun$ and $L=1\,L_{\odot}$ are adopted, using equations~(\ref{eq:omegap}), (\ref{eq:nebular-temp}), and (\ref{eq:nebular-cs}), the scale height $h$ can be described as
\begin{equation}
h = 5.0\times 10^{11} \left( \dfrac{\ro}{1 {\rm AU}} \right)^{5/4}  \ \ \ {\rm cm}.
\end{equation}

\subsection{Scaling}
Our basic equations can be normalized by unit time, $\Omega_{\rm p}^{-1}$  and unit length, $\hs$.
The density is also scalable in equations~(1) and (2) since we neglect
  the self-gravity of the disk in equation~(2).
We normalize the density by $\sigma_0/h$.
Hereafter the normalized quantities are expressed with tilde on top,
  e.g., $\tilde{x} = x/h$, $\tilde{\rho} = \rho/(\sigma_0/h)$, $\tilde{t}=t\,\Omega_{\rm p}$, etc. 
The non-dimensional unperturbed velocity and density are given by
\begin{equation}
 \vect{\tilde{v}} = ( 0, -\dfrac{3}{2}\tilde{x}, 0),
\end{equation}
\begin{equation}
 \tilde{\rho}_0 = 
  \dfrac{1}{\sqrt{2\pi}} {\rm exp} \left( - \dfrac{\tilde{z}^2}{2}\right).
\end{equation}
Thus, non-dimensional equations corresponding to
 equations~(\ref{eq:basic-1}), (\ref{eq:basic-2}), (\ref{eq:basic-3}),
 and (\ref{eq:phi-eff}) are 
\begin{equation}
 \dfrac{\partial \tilde{\rho}}{\partial \tilde{t}} + 
  \tilde{\nabla} \cdot (\tilde{\rho} \, \tilde{\vect{v}}) = 0,
\end{equation}
\begin{equation}
 \dfrac{\partial \tilde{\vect{v}}}{\partial \tilde{t}} + 
  (\tilde{\vect{v}} \cdot \tilde{\nabla}) \tilde{\vect{v}} =
  - \dfrac{1}{\tilde{\rho}} \tilde{\nabla} \tilde{P} - 
 \tilde{\nabla} \tilde{\Phi}_{\rm eff} - 2 \vect{\tilde{z}}  
\times \tilde{\vect{v}},  
\end{equation}
\begin{equation}
\tilde{P} = \tilde{\rho},
\end{equation}
\begin{equation}
\tilde{\Phi}_{\rm eff} = 
 - \dfrac{1}{2}(3 \tilde{x}^2 - \tilde{z}^2) \, - \, \dfrac{3{\rht}^3}{\tilde{r}},
\end{equation}
 where $\tilde{\vect{z}}$ is a unit vector directed to the $z$-axis.
Thus, the gas flow is characterized by only one parameter, the
 non-dimensional Hill radius $\rht=\rh/h$.
In this paper, we adopt $\rht =0.05-4.21 $ (see, Table~1).
As functions of  the orbital radius and the mass of
 the central star, the parameter $\rht$ are related to the actual mass of
 protoplanet in the unit of Jovian mass $M_{\rm J}$ as 
\begin{equation}
 \dfrac{M_{\rm p}}{M_{\rm J}} = 
 0.12 \left( \dfrac{M_{\rm c}}{1\msun} \right)^{-1/2} 
 \left( \dfrac{\ro}{1\,{\rm AU}} \right)^{3/4} \, \rht^3.
\end{equation}
For example, in the model with $\rht = 1.0$,  $\ro=5.2$\,AU and
 $M_{\rm c} = 1\,\msun$, the protoplanet mass is 
 $M_{\rm p} = 0.4 M_{\rm J}$ (model M04 in Table~\ref{table:table1}). 
Hereafter, we call model M04 `the fiducial model.'
For each model, the protoplanet mass for $\ro=5.2$\,AU and 
 $M_{\rm c} = 1\,\msun$ is presented in Table~\ref{table:table1}.  
In our parameter range,  at Jovian orbit ($\ro=5.2$\,AU), protoplanets have masses of $0.05-30\mj$.
We will show our results assuming $\ro = 5.2$ AU and 
 $M_{\rm c} = 1\msun$  in the following. 
We will discuss the dependence on the orbital radius $\ro$ in
 \S\ref{sec:jupiter-saturn}.

In subsequent sections, we use non-dimensional quantities  (e.g., $\tl{\rho}$, $\tl{x}$, $\tl{y}$, and $\tl{z}$) when we show the structure of the protoplanet system (Figs.~1, 4, 5, 7, 8, 12, and 13).
On the other hand, to compare physical quantities derived from numerical results with those of the present Jupiter, we use dimensional quantities at Jovian orbit when we show the time evolution or radial distribution (Figs.~2, 3, 6, 9, 10, and 11) of the mass $M$ and  (specific) angular momentum $J$ ($j$) of the protoplanet system, in which the conversion coefficient from physical quantities at Jovian orbit (5.2\,AU) into those at any orbit ($\ro$) is described in each axis.
In addition, for convenience, we redefine the time unit as $\deft \equiv \tl{t}/(2\pi) =1/(2\pi \Omega_{\rm p} )$ that corresponds to the orbital period of the protoplanet.
When $\ro=5.2$\,AU is assumed, $\deft=1$ corresponds to $11.86$\,yr.

\section{NUMERICAL METHOD}
\subsection{Nested-Grid Method}
\label{sec:nested-grid}
To estimate the angular momentum acquired by a protoplanet system from the protoplanetary disk, we need to cover a large dynamic range of spatial scale.
Using the nested-grid method \citep[for details, see][]{machida05a, machida06a}, we cover the region near the protoplanet by the grids with high spatial resolution, and region remote from the protoplanet by the grids with coarse spatial resolution. 
Each level of rectangular grid has the same number of cells ($ = 32 \times 128 \times 8 $), but cell width $\Delta \tilde{s}(l)$ depends on the grid level $l$. 
The cell width is reduced 1/2 with increasing the grid level ($l \rightarrow l+1$).
In a fiducial model, we use 8 grid levels ($l_{\rm max}=8$).
The box size of the  coarsest grid $l=1$ is chosen to $(\tl{L}_x, \tl{L}_y, \tl{L}_z) = (30, 120, 7.5)$, and that of the finest grid $l=8$ is $ (\tl{L}_x, \tl{L}_y, \tl{L}_z) = (0.234, 0.938, 0.059)$. 
The cell width in the coarsest grid $l=1$ is $\Delta \tilde{s} = 0.9375$, and it decreases with $\Delta \tilde{s}=0.9375/2^{l-1}$ as the grid level $l$ increases.
Thus, the finest grid has  $\Delta \tilde{s}(8)\simeq 7\times10^{-3}$.
We assume the fixed boundary condition in the $\tilde{x}$- and $\tilde{z}$-direction and the periodic boundary condition in the $\tilde{y}$-direction.

In real units, using the standard solar nebular model, the scale height at Jovian orbit ($\ro=$5.2\,AU) is $\hs=0.27$\,AU.
The computational domain in azimuthal direction corresponds to  $120\times0.27$\,AU =$32.4$\,AU, which is equivalent to the circumference of Jovian orbit around the Sun $2\pi\ro$ ($32.7$\,AU).
Although we imposed a periodic boundary condition in azimuthal direction, this domain size is valid except for ignoring the curvature.
Note that the computational domain is not necessarily the same as a real circumference of a planet as long as it is sufficiently large.
To verify our results, in some models, we calculate the evolution of the protoplanet system adopting different levels of the finest grid (or different maximum grid levels), $l_{\rm max}$ = 5, 6, 7, 9, and 10.
In these models, the cell width of $l_{\rm max} =5$ is $\Delta \tilde{s}= 0.23$ ($9.375\times 10^{13}\cm$ at $\ro=5.2$\,AU) that corresponds to 33 times Jovian radius, while that of $l_{\rm max}=10$ is $\Delta \tilde{s} = 1.83\times 10^{-3}$ ($7.32\times 10^9\cm$ at $\ro=5.2$\,AU) that corresponds to 1.02 time Jovian radius.
The maximum grid level $l_{\rm max}$, and the cell width of the finest grid for each model are summarized in Table~\ref{table:table1}.

\subsection{Test Simulation}
\label{sec:test}

\begin{figure}
\begin{center}
\includegraphics[width=130mm]{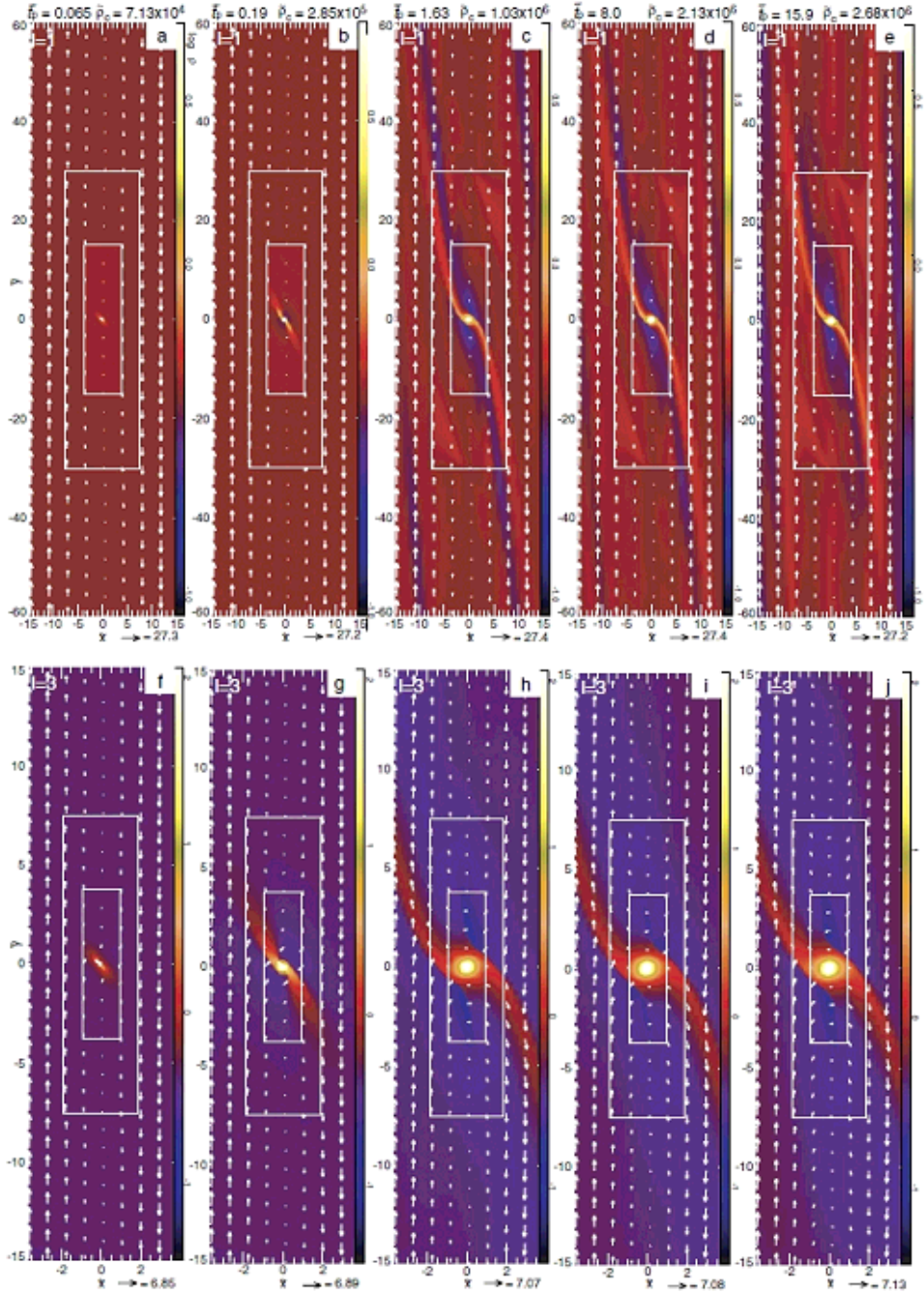}
\caption{
Time sequence for model M04.
Density ({\it color scale}) and velocity distribution ({\it arrows}) on the cross section in the $\tl{z}=0$ plane are plotted.
Each lower panel ($l=3$) is four times spatial magnification of each upper panel ($l=1$).
Three levels of grids are shown in each upper ($l=$1, 2, 3) and lower ($l=$3, 4, 5) panel.
Level of outermost grid is denoted in the upper left corner.
The elapsed time $\deft$ and central density $\tl{\rho}_{\rm c}$ on the midplane are denoted in the top of each panel.
The velocity scale in the unit of the sound speed is denoted in the bottom of each panel.
}
\label{fig:1}
\end{center}
\end{figure}

At first, we show the evolution of the protoplanet system calculated with  $l_{\rm max}=8$.
The cell width of the maximum grid level ($l=8$) that covers the region in the proximity of the protoplanet is $\Delta \tilde{s} = 7.3\times10^{-3}$\, ($\Delta s = 2.92\times 10^{10}$cm at $\ro=5.2$\,AU).
Figure~\ref{fig:1} shows the evolution for model MN04 (fiducial model), in which the protoplanet with $0.4\mj$ is adopted for $\ro=5.2$\,AU.
The upper panels in Figure~\ref{fig:1} (\ref{fig:1}{\it a} -- {\it e}) show the time sequence of the region far from the protoplanet ($\tilde{r} \gtrsim 15$) with  $l=$ 1, 2, and 3 grid levels, in which three different grid levels are superimposed, while the lower panels (Fig.~\ref{fig:1}{\it f}--{\it j}) show the region near the protoplanet ($\tilde{r}\lesssim 3.5$) with  $l=$ 3, 4, and 5.
Each lower panel is 4 times magnification of each upper panel.
In these panels, the protoplanet is located at the origin ($\tl{x}$, $\tl{y}$, $\tl{z}$) = (0, 0, 0). 
The elapsed time $\deft$ is denoted in each upper panel.
The central density $\tilde{\rho}_{\rm c}$ is also denoted in each upper panel.

Figures~\ref{fig:1}{\it a}--{\it c} and \ref{fig:1}{\it f}--{\it h} show that the density is enhanced in the narrow band with the spiral pattern that is distributed from the upper-left to the lower-right region.
The density gaps that appears on the right (left) side of the spiral pattern in the region of  $\tl{y}>0$ ($\tl{y}<0$) are also seen in these panels.
In Figure~\ref{fig:1}{\it c}, the density of the spiral pattern around the protoplanet is $\tl{\rho}\simeq2$, while that  of the gap is $\tl{\rho}\simeq 0.2$.
Thus, there are a density contrast of $\sim10$ between the spiral patterns and gaps.
In addition, the central density increases up to $\tl{\rho}\sim10^6$ for $\deft\gtrsim1$.
Figures~\ref{fig:1}{\it h}--{\it j} show a round shape near the protoplanet.
Since the Hill radius is $\rht = 1 $ in this model, the gravity of the protoplanet is predominant in the region of $\tilde{r} \ll1$.
Thus, the central region of $\tilde{r}\ll1$ has the round structure.
Figure~\ref{fig:1} shows that the density distributions in panels {\it c}--{\it e} (or {\it h}--{\it j}) are similar.
Except for the calculation with $l_{\rm max}=10$,  in all models, the structure around the protoplanet hardly changes for $\deft\gtrsim1$, which seems that the steady state is already achieved for $\deft\gtrsim1$.
\citet{tanigawa02a} investigated the evolution of the protoplanet system as the same condition as ours but in two dimensions, and showed that the gas stream around the protoplanet is in a steady state after a short timescale of $\deft \sim1$.

\subsection{Convergence Test}
In \S\ref{sec:test}, we showed the evolution of the protoplanet system with the maximum grid level $l_{\rm max}=8$.
In this subsection, to check the convergence of our calculation, we compare the evolutions of the protoplanet system with different maximum grid levels (or different cell width of the finest grid).
Since our purpose is to investigate the angular momentum acquired by the protoplanet system, we use the average specific angular momentum as a measure of the convergence.
As a function of the distance from the protoplanet $\tl{r}$, we define the average specific angular momentum $\tl{j}_r$  as 
\begin{equation}
\tl{j}_r = \dfrac{\tl{J}_r}{\tl{M}_r}, 
\label{eq:jr}
\end{equation}
where the mass 
\begin{equation}
\tl{M}_r = \int^{\tl{r}}_0  \, 4\pi \tl{r}^2 \tl{\rho}\, d\tl{r},
\label{eq:mr}
\end{equation}
and angular momentum 
\begin{equation}
\tl{J}_r  = \int^{\tl{r}}_0  4\pi \tl{r}^2 \tl{\rho} \, \tl{\varpi} \tl{v}_{\phi}\,  d\tl{r},
\label{eq:Jr}
\end{equation}
are integrated from the center $\tl{r}=0$ to distance $\tl{r}$.
Here, we adopt $\tl{r} = 0.5\, \rht$ (i.e., $\tl{j}_{0.5\rht}$).
The dependence of $\tl{j}_r$ on $\tl{r}$ is discussed in \S\ref{sec:dis-evo}.
We often show the mass  and (specific) angular momentum of the protoplanet system in real units at $\ro=5.2$\,AU to compare numerical results with present values of gaseous planets.
The dimensional values $M_r (5.2\,{\rm AU})$, $J_r (5.2\,{\rm AU})$, and  $j_r (5.2\,{\rm AU})$ at $\ro=5.2$\,AU can be converted into  $M_r (\ro)$, $J_r (\ro)$, and  $j_r (\ro)$ at any orbit as
\begin{equation}
M_r (\ro) = M_r (5.2\,{\rm AU}) \left( \dfrac{\ro}{{\rm 5.2AU}} \right),
\end{equation}
\begin{equation}
J_r (\ro) = J_r (5.2\,{\rm AU}) \left( \dfrac{\ro}{{\rm 5.2AU}} \right)^{-7/4},
\end{equation}
and
\begin{equation}
j_r (\ro) = j_r (5.2\,{\rm AU}) \left( \dfrac{\ro}{{\rm 5.2AU}} \right).
\end{equation}
In the following, we show  the values at 5.2\,AU.
When we refer to dimensional physical quantities without any mention of orbit, they are the values at $\ro = 5.2$ AU.

\begin{figure}
\begin{center}
\includegraphics[width=150mm]{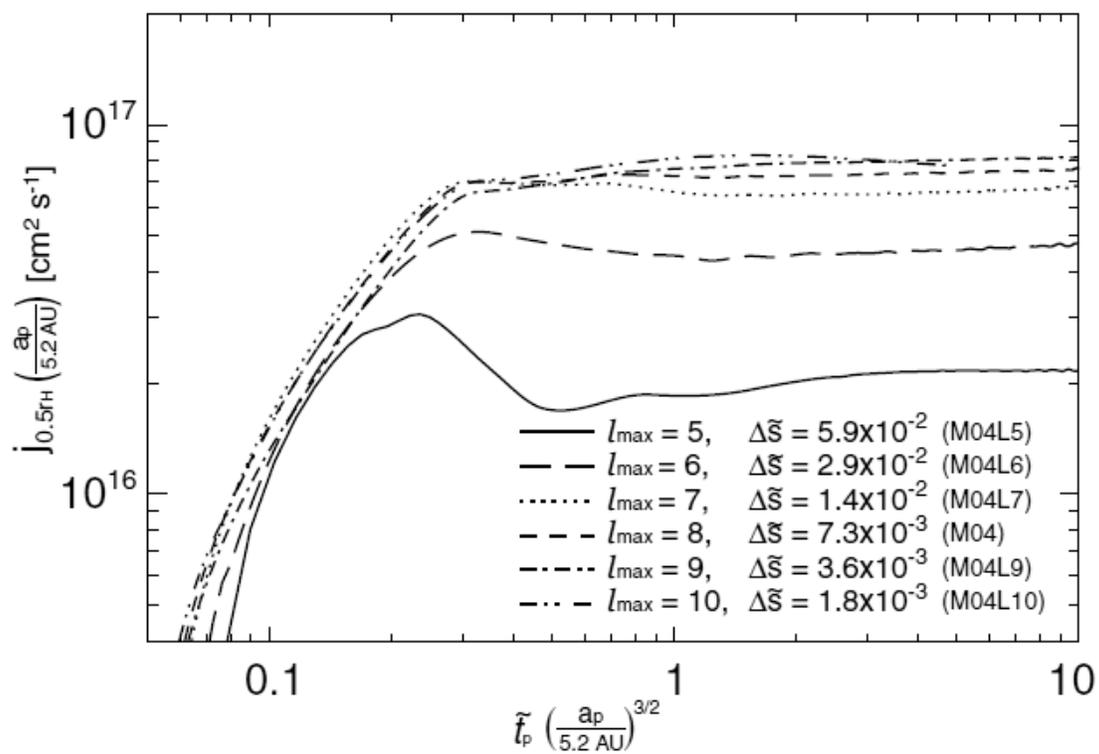}
\caption{
Evolution of average specific angular momenta derived in the region of $r<0.5\,\rh$ for models with different finest grid levels $l_{\rm max}$.
}
\label{fig:2}
\end{center}
\end{figure}

The evolution of  $j_{0.5\rh}$ for models M04L5--M04L9 (see, Table~\ref{table:table1}) are shown in Figure~\ref{fig:2}.
In these models, the protoplanet mass is fixed, and only the maximum grid level (or cell width of the finest grid) is changed.
Figure~\ref{fig:2} shows that the average specific angular momentum rapidly increases initially, then it saturates at a certain value in each model.
Although we adopted the same mass of the protoplanet, the saturation levels of the average specific angular momenta are different.
The average specific angular momentum saturates at $j_{0.5\rh} \simeq 2\times 10^{16}\jcm$ for model M04L5, while it is saturated at $j_{0.5\rh} \simeq 7-8\times 10^{16}\jcm$ for models M04L7, M04, M04L9, and M04L10.
Thus, there are a little differences for models with $l_{\rm max} > 7$.
For example, model M04L7 has $j_{0.5\rh} = 6.9\times 10^{16}\jcm$, while model M04L9 has $j_{0.5\rh} = 8.2\times 10^{16}\jcm$ at $\deft=10$.
Therefore, the average specific angular momenta are sufficiently converged in $l_{\rm max}>7$ or $\Delta \tl{s} < 1.4\times 10^{-2}$ within a relative error of about 15\%.
In the following, we safely calculate the evolution of the protoplanet system with $l_{\rm max}=8$ of the maximum grid level.

\subsection{Models with and without Sink Cells}
\label{sec:sink}
When we adopt the maximum grid level $l_{\rm max}=8$, the cell width is $\Delta \tl{s} = 7.3\times 10^{-3}$.
In real units, when the protoplanet is located at 5.2\,AU, the cell width corresponds to $\Delta s = 2.9\times 10^{10}$cm whose size is 4.1 times the Jovian radius.
Interior to the gas giant planet, there exists a solid core with $\simeq 10\me$ in mass and  $\simeq10^9$\,cm in size \citep{mizuno80}. 
The size of the solid core is much smaller than the cell width adopted in our calculation.
To investigate the evolution of a protoplanet through gas accretion, in principle, we need to resolve a solid core with sufficiently small cell size ($\Delta s \ll 10^9$ cm).
However, since our purpose is to investigate the angular momentum flowing into the protoplanet system, we do not always need to resolve a central solid core and a protoplanet.
As discussed in \citet{dangelo02}, the gas around the central region makes an artificial pressure gradient force, which may  affect the gas accretion onto the protoplanet system.
To check this, we also calculated the evolution of the protoplanet system adopting the sink cell in some models.
We parameterized the size of the sink: $\tl{r}_{\rm sink} =  0.01$ and $0.03$ (models M04S01 and  M04S03).
These models are also summarized in Table~\ref{table:table1}.

In real units, the radius of the sink in model M04S01 is $4.0\times 10^{10}\,{\rm cm}\,$ ($\simeq 5.6$ Jovian radius), while that in model M04S03 is $1.2\times 10^{11}\, {\rm cm}\,$( $\simeq 17$ Jovian radius).
During the calculation, we remove the gas from the region inside the sink radius in each time step, and integrate the removed mass and angular momentum that are assumed as the mass and angular momentum of the protoplanet system.

\begin{figure}
\begin{center}
\includegraphics[width=150mm]{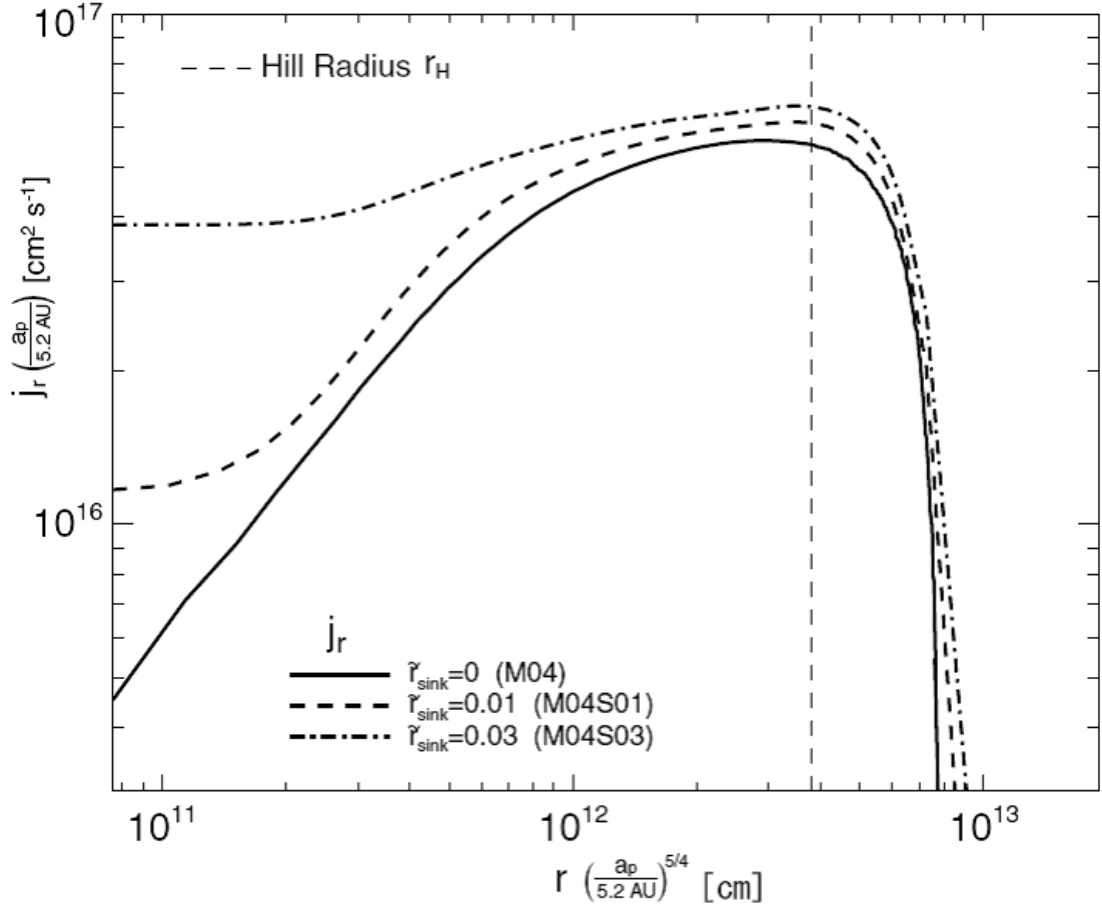}
\caption{
The average specific angular momenta $j_r$ against the distance from the origin with models having different sink radii  $\tl{r}_{\rm sink}$.
The vertical dotted line represents the Hill radius.
}
\label{fig:3}
\end{center}
\end{figure}

After the steady states are achieved at $\deft\simeq 20$, we estimate the average specific angular momenta $\tl{j}_r$ as a function of $\tl{r}$.
Figure~\ref{fig:3} shows the distribution of the average specific angular momentum  $j_r$ against the distance from the protoplanet $r$ in real units for each model.
The solid line represents $j_r$ without a sink cell, while the other lines represent those with sink cells.
Inside the sink radius, the average specific angular momentum is assumed as a constant value.
In Figure~\ref{fig:3}, the vertical dotted line indicates the Hill radius.

In Figure~\ref{fig:3} the solid line shows that  $j_r$ increases from the center to a peak around the Hill radius, and drops sharply just outside the Hill radius.
The drop indicates that the angular momentum becomes negative at $r>\rh$.
Thus, the rotational direction turns around between the region inside and outside the Hill radius.
As shown in \citet{sekiya87}, \citet{miyoshi99}, and \citet{tanigawa02a}, the protoplanet formed by the gas accretion in the protoplanetary disk has a prograde spin, and thus it has a positive (specific) angular momentum.
On the other hand, gas far outside of Hill sphere seems to rotate retrogradely against the protoplanet because it rotates with nearly Keplerian velocity with respect to the central star [$v=-(3/2)\, \Omega_{\rm p}\,x$,  as shown in eq.~(\ref{eq:shear})].
As a result, gas inside the Hill radius has a positive angular momentum, while that outside the Hill radius has a negative angular momentum.

Figure~\ref{fig:3} shows that $j_r$ in models with the sink cell depends on the radius of the sink in the region $r \ll\rh$.
However, they do not depend sensitively on the size of the sink in the region of $r \gtrsim 0.5\,\rh$.
For example, at $r=\rh$, model without the sink (model M04) has $j_{\rm \rh}=5.3\times 10^{16}\jcm$, while model M04S03 has $j_{\rm \rh} = 6.4\times 10^{16}\jcm$.
Thus, difference of the average specific angular momentum among these models at the Hill radius is $\simeq12$\%.
This difference decreases with the sink radius.
In this study, we focus on the angular momentum of the protoplanet system, not the planet itself.
As shown in Figure~\ref{fig:3}, the angular momentum acquired by the protoplanet system (planet + circumplanetary disk) extends up to the Hill radius, in which almost all the angular momentum distributed in the region of $r \gg \rp$ or $r \gg r_{\rm sink}$.
In the following, we calculate the evolution of the protoplanet system without the sink cell.

\section{RESULTS}
\subsection{Typical Gas Flow}
\label{sec:typical}
\begin{figure}
\begin{center}
\includegraphics[width=150mm]{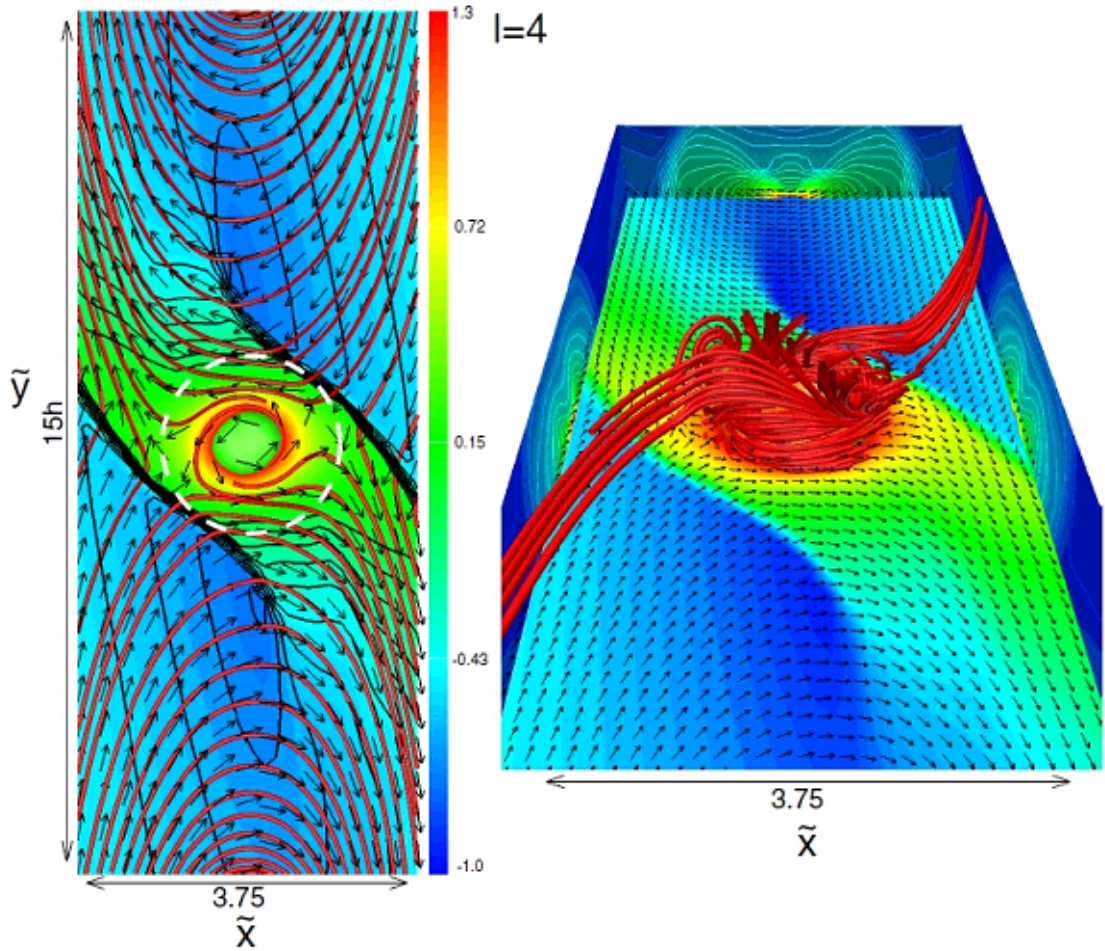}
\caption{
Structure around the Hill sphere for model M04 on the midplane ({\it left}) and  in three dimension shown in bird's-eye view ({\it right}).
The gas streamlines ({\it red lines}), density structure ({\it color}), and velocity vectors ({\it arrows}) are plotted in each panel.
The dotted circle in the left panel represents the Hill radius.
The size of the domain is shown in each panel.
}
\label{fig:4}
\end{center}
\end{figure}

Figure~\ref{fig:4} shows the gas structure around the Hill sphere for model M04 at $\deft=20$.
Figure~\ref{fig:4} left panel shows the structure on the cross section in the $\tl{z}=0$ plane, in which red lines indicate the streamlines.
In this panel, the gas enters $l=4$ grid from upper $\tl{y}$ boundary for $\tl{x} > 0$ (from lower $\tl{y}$ boundary for $\tl{x} < 0$) and goes downward (upward for $\tl{x} < 0$) according to the Keplerian shear motion.
The shocks (crowded contours near the Hill radius) are seen in the upper right and lower left region from the protoplanet.
In this model, the shock front almost corresponds to the Hill radius (Fig.~\ref{fig:4} left panel).
When the gas approaches the protoplanet, the streamlines are bent by the gravity of the protoplanet.
According to \citet{miyoshi99}, the gas flow is divided into three region: the pass-by region ($|\tl{x}|\gtrsim \rht$), the horseshoe region ($\tl{x}\lesssim \rht$, and $\vert \tl{y} \vert \gtrsim \rht$), and the planet atmosphere region ($\tl{r} \lesssim \rht$).
Note that although \citet{miyoshi99} classified the flow pattern in their two dimensional calculation, their classification is useful for global flow pattern in three dimensions.
In the pass-by region, the flow is first attracted toward the protoplanet, and then causes a shock after passing by the protoplanet.
At the shock front, the density reaches a local peak and the streamlines bend suddenly.
On the other hand, the gas entering the horseshoe region turns round by the Coriolis force and goes back.
The outermost streamlines in the horseshoe region (i.e., the streamlines passing very close to the protoplanet) pass through the shock front, while the gas on the streamlines far from the protoplanet does not experience the shock.
In the atmospheric region, the gas is bound by the protoplanet and forms a circumplanetary disk that revolves circularly around the protoplanet in the prograde (counterclockwise) direction.

Although the streamlines on $\tl{z}=0$ plane (Fig.~\ref{fig:4} left panel) are similar to those in recent two dimensional calculations \citep[e.g.,][]{lubow99,tanigawa02a}, there are important differences.
In two-dimensional calculations, a part of the gas near the Hill sphere can accrete onto the protoplanet.
\citet{lubow99} showed that gas only in a narrow band that distributed from the lower left to the upper right region against the protoplanet for $\tl{y}<0$ spirals inward toward the protoplanet passing through the shocked region and finally accretes onto the protoplanet \citep[for details, see Figs.~4, and 8 of][]{lubow99}.
On the other hand, in our three-dimensional calculation,  gas only flows out from the Hill sphere and thus does not accrete onto the protoplanet on the midplane.
Figure~\ref{fig:4} left panel shows that although gas flows into the Hill sphere, a part of the gas flows out from the central region.
Figure~\ref{fig:4} right panel is three-dimensional view at the same epoch as the left panel.
In this panel, only the streamlines flowing into the high-density region of $\tl{r} \ll \rht$ are drawn for $\tl{z} \ge 0$ which  are inversely integrated from the high-density region.
This panel clearly shows the gas flowing into the protoplanet in the vertical direction.

\begin{figure}
\begin{center}
\includegraphics[width=150mm]{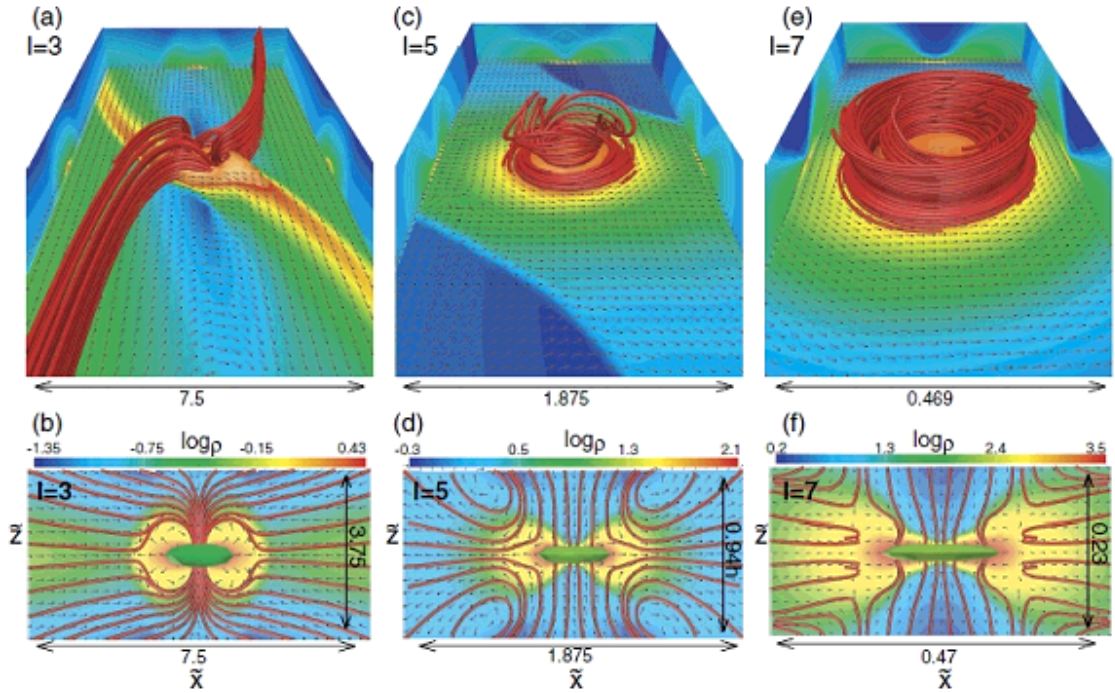}
\caption{
Structure around the protoplanet for model M04 with different grid levels: $l=3$ ({\it left}), 5 ({\it middle}), and 7 ({\it right}).
Each upper panel is the structure in three dimensions shown in bird's-eye view ({\it right}), and each lower panel is the structure on the cross section in the $\tl{y}=0$  plane.
The gas streamlines ({\it red lines}), density structure ({\it color}), and velocity vectors ({\it arrows}) are plotted in each panel.
The size of the domain is shown  in each panel.
}
\label{fig:5}
\end{center}
\end{figure}

To investigate gas flowing into the protoplanet system in detail,  in Figure~\ref{fig:5}, we plot the streamlines at the same epoch as Figure~\ref{fig:4} with different grid levels ($l$=3, 5, and 7).
In this figure, each upper panel shows three-dimensional view, while each lower panel shows the structure on the cross section in the $\tl{y}=0$ plane.
Note that, in lower panels, the streamlines are projected onto the $\tl{y}=0$ plane.
Figure~\ref{fig:5}{\it a} shows only the streamlines in a narrow bundle flowing into the protoplanet system.
This feature is similar to that shown in two-dimensional calculations \citep{lubow99, tanigawa02a}.
However, the streamlines in Figure~\ref{fig:5}{\it a} indicate that gas rises upward near the shock front and then falls into the central region in the vertical direction.
Gas flowing in the vertical direction spirals into the inner region (Fig.~\ref{fig:5}{\it c}).
In this process, vortices appear as shown in Figure~\ref{fig:5}{\it d}.
As shown in Figure~\ref{fig:4} left panel, also in Figure~\ref{fig:5}{\it d}, gas is flowing out from the central region on $\tl{z}=0$ plane.
Gas in the proximity of the protoplanet rotates circularly in the prograde direction as shown in Figure~\ref{fig:5}{\it e}.
Figure~\ref{fig:5}{\it f} shows that a part of the gas  flowing into the upper boundary of $l=7$ grid level contributes to the disk formation around the protoplanet, while a remainder is bent and flows out from the central region.

When we look down the protoplanetary disk from the above along the $\tl{z}$-axis, streamlines may seem to be almost the same as those in two-dimensional calculations.
However, gas moves also in the vertical direction: streamlines go upward near the shock front ($\tl{r} \sim \rht$), and vertically falls into the central region at $\tl{r} \ll \rht$.
This feature of streamlines is also seen in models M04S01 and  M04S03, in which the sink cell is adopted.
Thus, different features of streamlines in two- and three-dimensional calculations are not caused by the effect of the pressure gradient force, but caused by the dimensions (because the same feature appears in both models with and without the sink).
This flow pattern is also seen in other three-dimensional calculation \citep{tanigawa02b,dangelo03}.
In two-dimensional calculation, since the vertical motion is restricted, the flow pattern is different from that in three dimensions.
This difference affects the accretion rate onto the protoplanet and migration rate.
\citet{dangelo03} showed that the migration rate is different between two- and three-dimensional calculations.
In the present study, however, since we focus on the angular momentum of a protoplanet system, we do not discuss them any more.
We will discuss the accretion and migration rate in the subsequent papers.

Finally, we comment on the circumplanetary disk.
In Figure~\ref{fig:5} lower panels, the green surface (i.e., iso-density surface) indicates the high-density structure around the protoplanet.
These panels show the disk-like structure in the proximity of the protoplanet, and the disk becomes thinner as it approaches the protoplanet (i.e., the origin).
We discuss the circumplanetary disk in \S\ref{sec:dis-disk}.

\subsection{Dependence on Protoplanet Mass}
\begin{figure}
\begin{center}
\includegraphics[width=160mm]{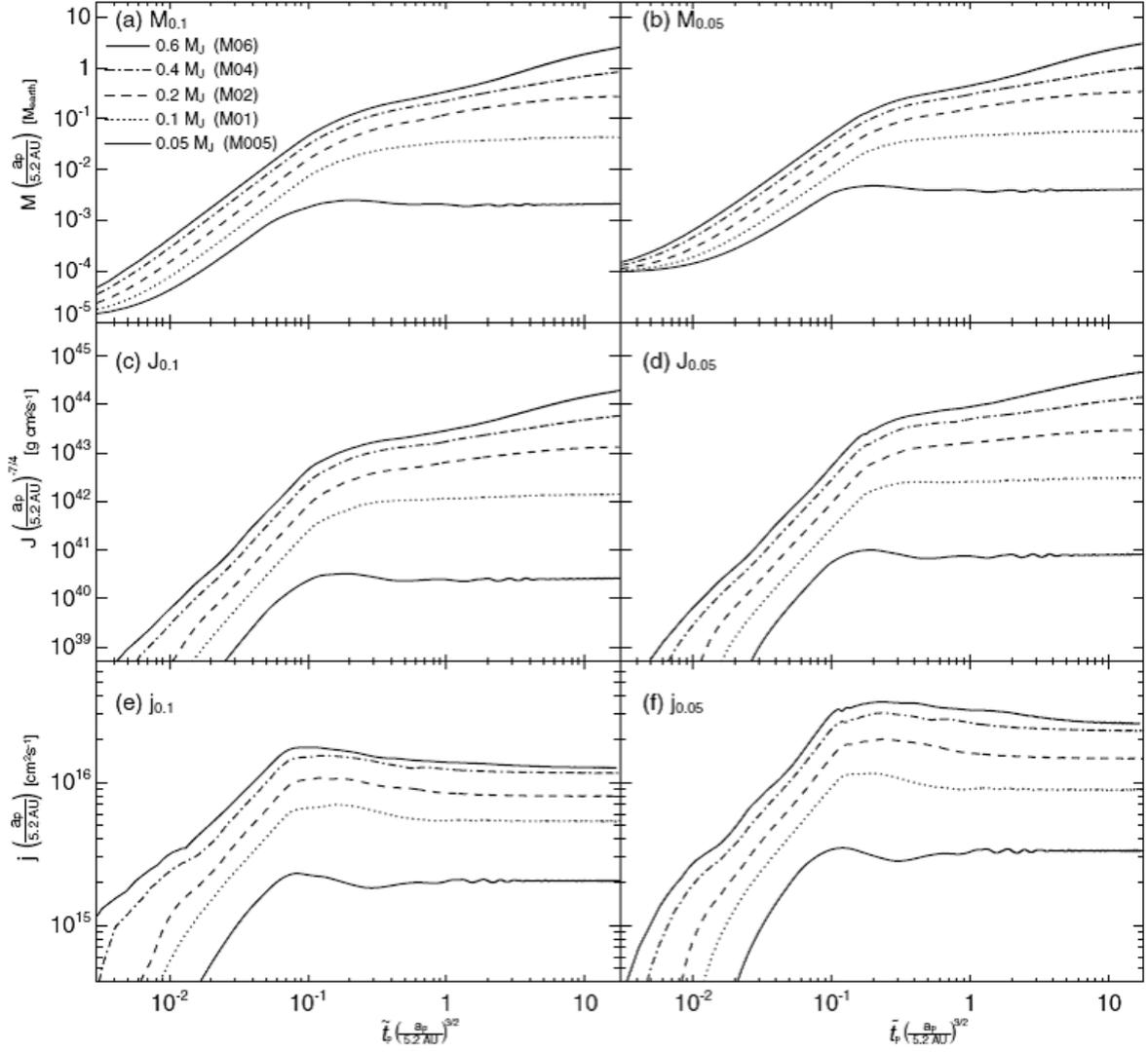}
\caption{
Evolution of accumulated masses (panels {\it a} and {\it b}), angular momenta (panels {\it c} and {\it d}), and the average specific angular momenta (panels {\it e} and {\it f}) in the region of $\tl{r}<0.1$ (panels {\it a}, {\it c}, and {\it e}) and $\tl{r}<0.05$ (panels {\it b}, {\it d}, and {\it f}) against the elapsed time for models M005, M01, M02, M04 and M06.
}
\label{fig:6}
\end{center}
\end{figure}

We have shown the evolution of the protoplanet system for model with 0.4$\mj$ in \S\ref{sec:typical}.
In this subsection, we investigate the evolution of the protoplanet system with different protoplanet masses.
Figure~\ref{fig:6} upper panels show the accumulated masses (eq.~\ref{eq:mr}) within $\tl{r}<0.1$ ($M_{\rm 0.1}$, Fig.~\ref{fig:6}{\it a}) and  $\tl{r}<0.05$ ($M_{\rm 0.05}$, Fig.~\ref{fig:6}{\it b}) against the elapsed time for different models, while Figure~\ref{fig:6} middle panels show the corresponding angular momenta (eq.~\ref{eq:Jr}) in the same regions  (Fig.~\ref{fig:6}{\it c} for $J_{0.1}$, and Fig.~\ref{fig:6}{\it d} for $J_{0.05}$).
In Figure~\ref{fig:6}, both masses and angular momenta for all models increase rapidly for $\deft<0.1$.
This is because the protoplanet with mass of $0.05-0.6\mj$ suddenly appears in the protoplanetary disk at $\deft=0$.
However, this rapid growth phase ($\deft<0.1$) is not real, because the gas is considered to begin to accrete onto the protoplanet when the mass of the solid core exceeds $\simeq 10\me$ \citep{mizuno80,bodenheimer86,ikoma00}.
The growth rates of the mass and angular momentum begin to decrease at $\deft \sim0.1$ in all models, then both masses and angular momenta increase with an almost constant rate until the end of the calculation ($0.1 \lesssim \deft \lesssim 20$).
\citet{tanigawa02a} calculated the mass accretion rate onto the protoplanet as
\begin{equation}
\dot{M} = 8.0\times 10^{-3} \me \left( \dfrac{M_{\rm p}}{10 \me} \right)^{1.3}. 
\end{equation}
Thus, the growth time $\tau_{\rm grow} = M/\dot{M}$ is  $\tau_{\rm grow} = 1000-430$\,yr (i.e., $\deft=88-36$).
In our calculation, we continue to calculate the evolution of the protoplanet system for $t\sim 230$\,yr ($\deft\sim20$) by fixing the planet mass.
We think that this treatment is not problematic, since the growth timescale is longer than our calculation time.
Note that since \citet{tanigawa02a} calculated the evolution of the protoplanet system in two dimensions, the growth rate might be different from that in  three-dimensional calculations.

Figure~\ref{fig:6} lower panels show the evolution of the average specific angular momentum in the region of $\tl{r}<0.1$ (Fig.~\ref{fig:6}{\it e} for $j_{0.1}$) and $\tl{r}<0.05$ (Fig.~\ref{fig:6}{\it f} for $j_{0.05}$).
The masses and angular momenta flowing into the protoplanet system increase with a constant rate for $\deft<0.1$, while the average specific angular momenta are saturated at certain values for $\deft>0.1$.
This saturation means that flow around the protoplanet is in the steady state.
Figures~\ref{fig:6}{\it e} and {\it f} also indicate that the average specific angular momentum brought into the protoplanet system increases with the mass of the protoplanet.
In Figures~\ref{fig:6}{\it e} and {\it f}, the average specific angular momenta in the region of $\tl{r}<0.1$ is larger than those in the region of $\tl{r}<0.05$ indicating  that the protoplanet system has a larger average specific angular momentum in the more distant place from the protoplanet.  
We will investigate the angular momentum acquired in the protoplanet system in \S\ref{sec:dis-evo}.

\subsection{Gas Structure around a Protoplanet}
\begin{figure}
\begin{center}
\includegraphics[width=150mm]{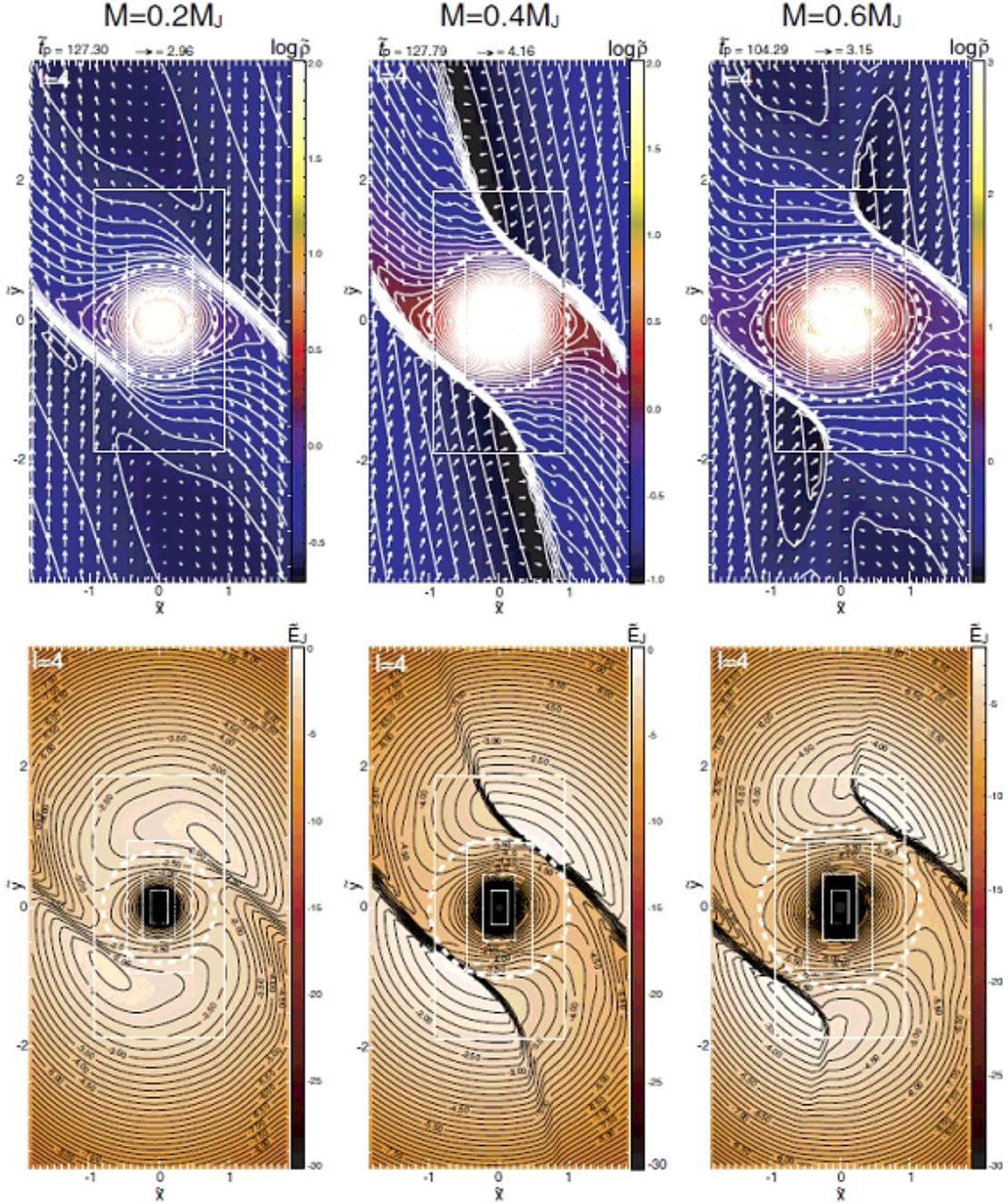}
\caption{
The density ({\it false color, {\rm and} contours}), and velocity vectors ({\it arrows}) are plotted on the cross section in the $\tl{z}=0$ plane in each upper panel for models M02 ({\it left}), M04 ({\it middle}), and M06 ({\it right}).
The Jacobi energy ({\it false color, {\rm and} contours}) are plotted in each lower panel for the same models as in each  upper panel.
Three levels of grid ($l=$4, 5, and 6) are superimposed in each panel.
The dotted-circle represents the Hill radius.
}
\label{fig:7}
\end{center}
\end{figure}

\begin{figure}
\begin{center}
\includegraphics[width=150mm]{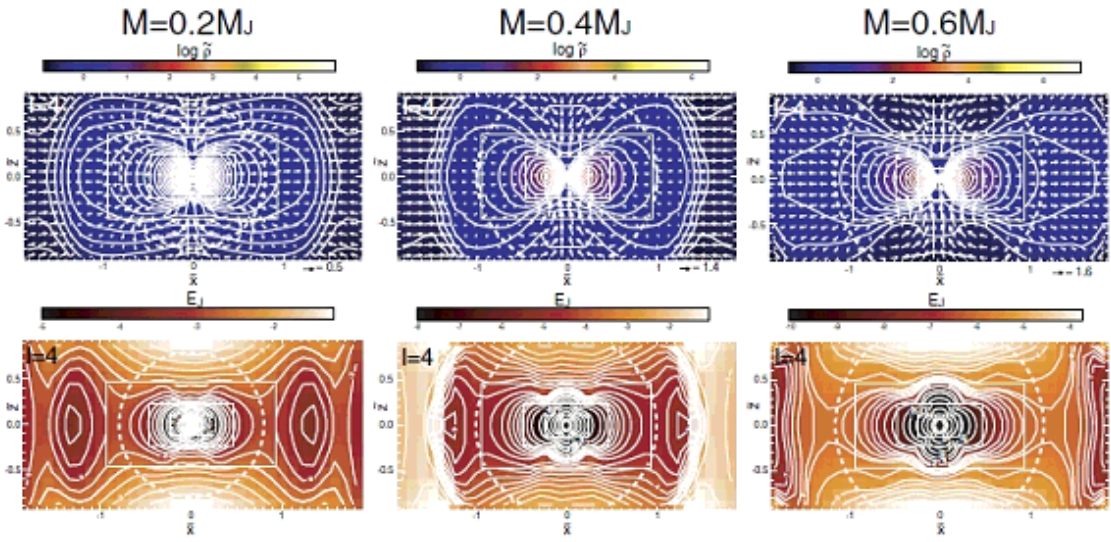}
\caption{
Same as Fig.~\ref{fig:7}, but in the cross section in the $\tl{y}=0$ plane.
}
\label{fig:8}
\end{center}
\end{figure}

Figures~\ref{fig:7} and \ref{fig:8} show the density distributions (upper panels) and Jacobi energy contours (lower panels)  on the cross section in the $\tl{z}=0$ (Fig.~\ref{fig:7}) and $\tl{y}=0$ plane (Fig.~\ref{fig:8}) around the Hill sphere after the steady state is achieved ($\deft\sim 20$) for models M02 (left panels), M04 (middle panels), and M06 (right panels).
The white-dotted line in each panel indicates the Hill radius $\rht$.
In each upper panel, the shock appears from the upper left to lower right near the Hill radius.
These shocks are frequently seen in similar calculations \citep{sekiya87,miyoshi99,lubow99,tanigawa02a,dangelo02}.
In Figure~\ref{fig:7} upper panels, the round structures are seen in the proximity to the protoplanet ($\tl{r} \ll \rht$), while the ellipsoidal structures are seen in the regions of $\tl{r}\simeq \rht$.
This is because gas distributed near the protoplanet is more strongly bound by the protoplanet.
Figure~\ref{fig:8} upper panels show that contours of the central region sags  in the center of a concave structure, and thin disks are formed around the protoplanet ($\tl{r} \ll \rht$).
In addition, the butterfly-like structure is also seen inside the Hill radius in Figure~\ref{fig:8} upper panels.
These structures are considered to be formed by the rapid rotation of the central circumplanetary disk: similar structure is seen in a rapidly rotating protostar \citep[e.g., Fig.1 of][]{saigo06}.
We will discuss the disk structure in \S\ref{sec:dis-disk}.

In contrast to celestial mechanics, it is difficult to find fluid elements bound by the protoplanet because thermal energy is important in addition to the gravitational and kinetic energies.
To discern gas bound by the gravity of the protoplanet, we use the Jacobi energy as an indicator \citep{canup95,kokubo00}.
In our unit, the Jacobi energy is given by
\begin{equation}
\tl{E}_{\rm J} = \dfrac{1}{2}\left(\dot{\tl{x}}^2 + \dot{\tl{y}}^2 + \dot{\tl{z}}^2\right) - \dfrac{3}{2}\tl{x}^2 +\dfrac{1}{2}\tl{z}^2  - \dfrac{3\rht^3}{\tl{r}}.
\label{eq:jacobi}
\end{equation} 
In equation~(\ref{eq:jacobi}), the first term is the kinetic energy, the second and third terms are the tidal energy of the central star, and the forth term is the gravitational energy of the protoplanet.
The Jacobi energy is the conserved quantity in a rotating system. 
In the case of fluid, the Jacobi energy is not strictly appropriate because the thermal energy is ignored.
However, we can use this for rough estimation of gas bound by the protoplanet.
We determine fluid elements bound by the protoplanet from the contour of the Jacobi energy in the lower panels in Figures~\ref{fig:7} and \ref{fig:8}.

Fluid elements with lower Jacobi energy are strongly bound by the protoplanet as described in equation~(\ref{eq:jacobi}).
Lower panels of Figures~\ref{fig:7} and \ref{fig:8} show that inside the region of $\tl{r} \lesssim 0.5\, \rht$, each contour has a closed ellipse.
Since fluid elements move on this closed orbit, it is considered that gas distributed in the region of $\tl{r} \lesssim 0.5\, \rht$ is bound by the protoplanet when the thermal effect can be ignored.
We discuss the thermal effect in \S\ref{sec:dis-evo}.
On the other hand, although fluid elements exist inside the Hill radius, outside $0.5\,\rht \lesssim \tl{r} \lesssim \rht$,  they are not bound by the protoplanet, because the contours of the Jacobi energy straddle the Hill radius.
For example, in lower left panel of Figure~\ref{fig:7}, the contour of $\tl{E}_{\rm J} = -2.5$ straddle the Hill radius, and thus fluid elements  with this Jacobi energy freely move on this contour.
Thus, although these elements transiently stay inside the Hill radius, they flow out from the Hill sphere.
In summary, Figures~\ref{fig:7} and \ref{fig:8} suggest that fluid elements  inside $\tl{r} \lesssim 0.5\,\rht$ are bound by the protoplanet.

\subsection{Angular Momentum of a Protoplanet System}
\label{sec:res-ang}
\begin{figure}
\begin{center}
\includegraphics[width=150mm]{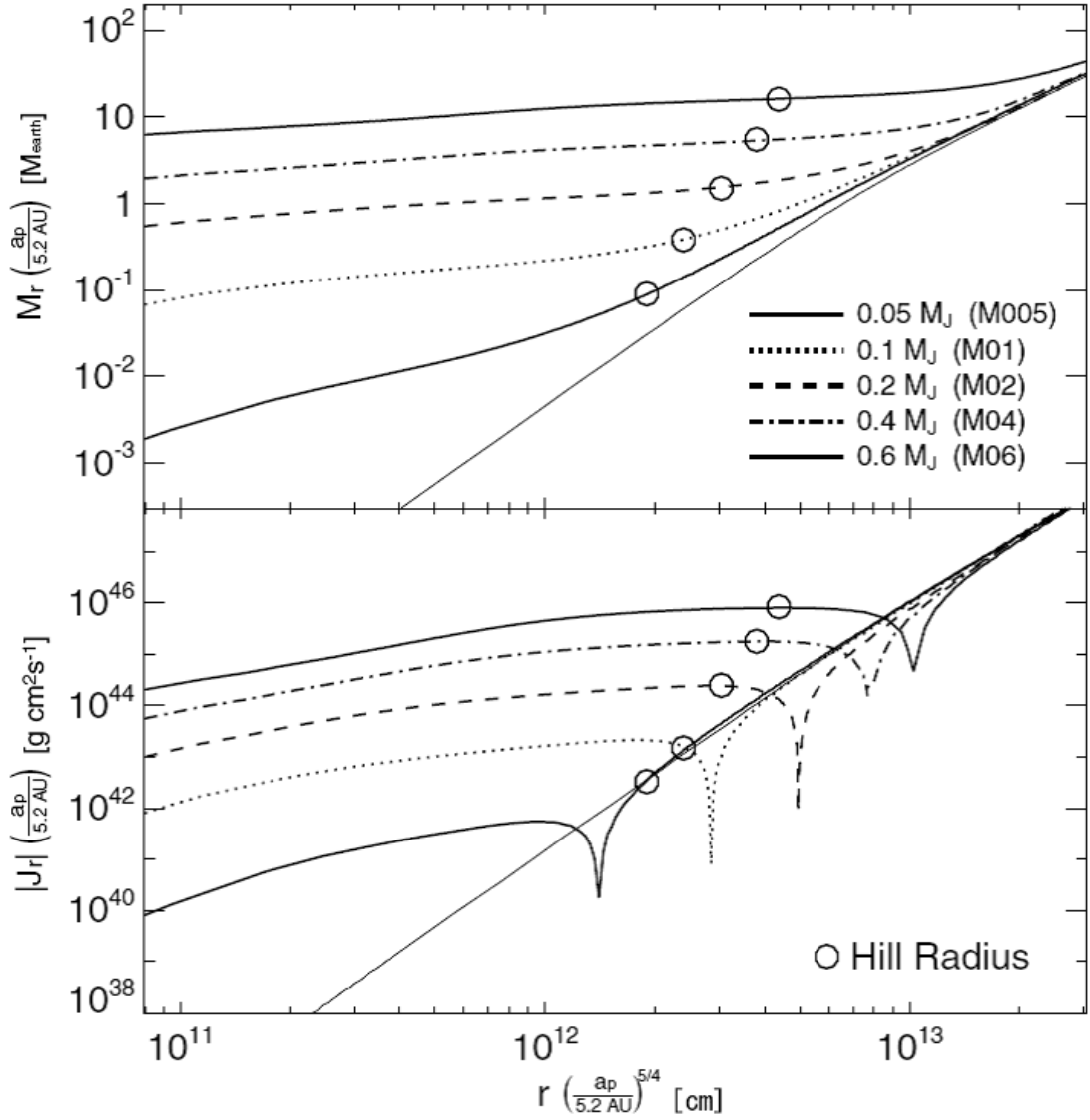}
\caption{
The distribution of accumulated mass (upper panel) and absolute value of the angular momentum (lower panel) against the distance from the protoplanet for models with M005, M01, M02, M04, and M06.
The thin solid lines represent the initial distribution of the mass (upper panel) and angular momentum (lower panel).
The circles represent the Hill radii.
}
\label{fig:9}
\end{center}
\end{figure}

Figure~\ref{fig:9} upper panel shows the accumulated mass $M_r$ for different models after the steady state is achieved ($\deft \simeq 20$).
In this panel, a thin solid line indicates the initial value (or the value of the protoplanetary disk), and the circle is the Hill radius $\rh$ for each model.
The accumulated mass in any model is larger than the initial value, because gas flows into the Hill sphere.
This panel indicates that the massive protoplanet has a massive envelope.
Since these mass distributions are in a steady state at the fixed mass of the protoplanet, it can be considered that different curves correspond to snapshots at different evolution phases.
Namely, a gas envelope increases its mass with time (or the protoplanet mass). 
Outside the Hill radius, the accumulated mass in each model converges to the initial value indicating that the mass distribution for $r \gg \rh$ does not change from the initial state because the influence of the protoplanet is small.

Figure~\ref{fig:9} lower panel shows the absolute value of the angular momentum $\vert J_r \vert $.
There are spikes in all models, at which the sign of the angular momentum is reversed.
As shown in \S\ref{sec:sink},  the angular momentum has a positive sign around the protoplanet ($r \ll \rh$), while it becomes  negative outside the Hill radius ($r \gg \rh$). 
The sign of the  angular momentum is reversed outside the Hill radius for models M01, M02, M04, and M06, while it is reversed inside the Hill radius for model M005. 
This is because the protoplanet system for model M005 does not acquire a sufficient mass and angular momentum owing to the shallow gravitational potential and relatively large thermal pressure (for details, see \S\ref{sec:dis-evo}).
Figure~\ref{fig:9} upper panel shows that the protoplanet system has the envelope mass of  only $M\simeq 0.06\me$ for model M005, in which the protoplanet mass is $5\mj$.
Except for model M005, the angular momenta gradually decrease after they reach their peak around $r \simeq \rh$, then it becomes negative at $r \simeq 1-2\rh$.  
Thus, the angular momenta bound by the protoplanet system are limited in the region of $r < 2\rh$ at the maximum.
Figure~\ref{fig:9} lower panel shows that the angular momentum keeps an almost constant value around the Hill radius ($0.5\,\rh \lesssim r \lesssim 1\, \rh$), until they are reversed.
This means that the angular momentum with plus sign and that with minus sign are mixed in this region, as shown in Figure~\ref{fig:4} left panel.
As a result, wherever we estimate the angular momentum in the range of $0.5\,\rh \lesssim r \lesssim 1\,\rh$, we can obtain almost the same values of the angular momentum.

\begin{figure}
\begin{center}
\includegraphics[width=150mm]{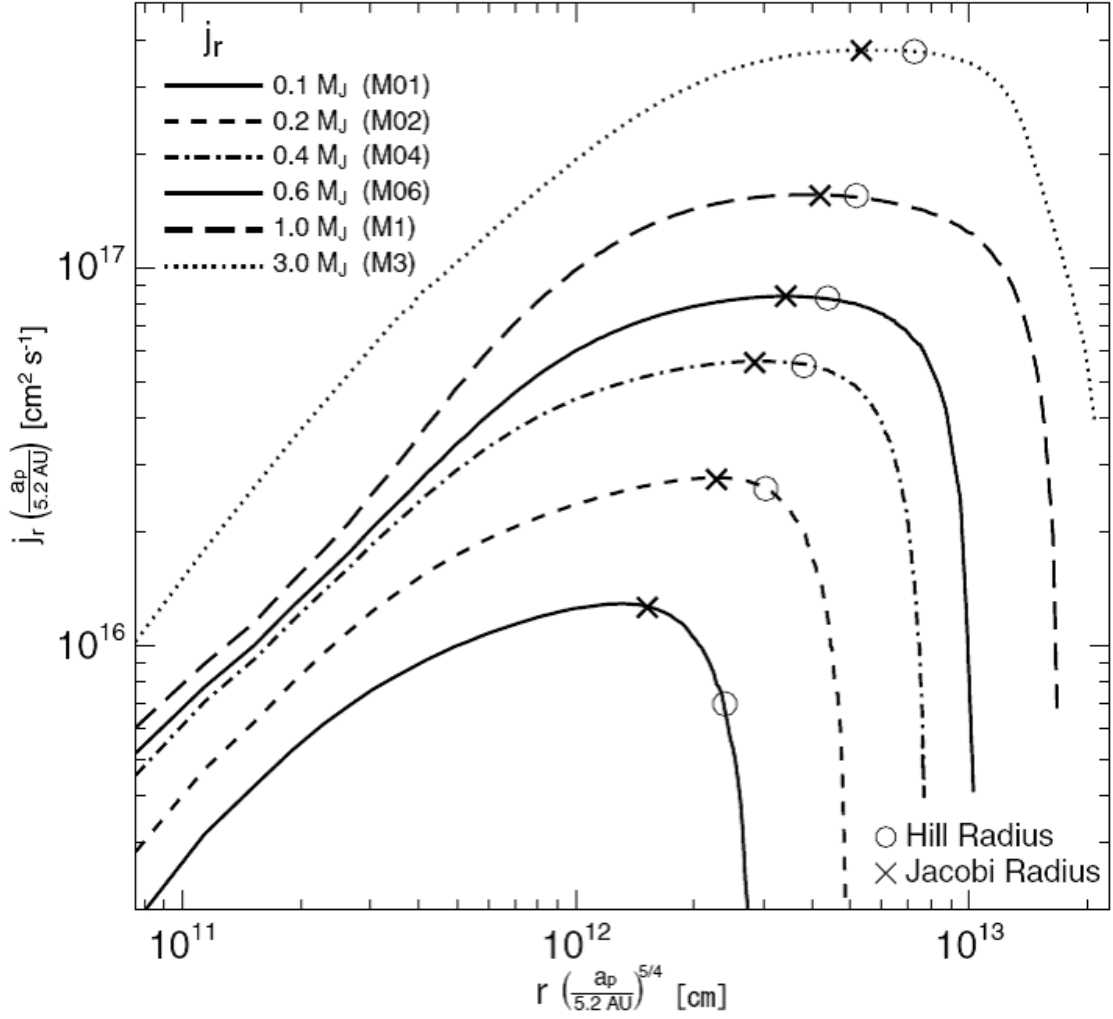}
\caption{
Average specific angular momentum $j_r$ against the distance from the protoplanet for models with M01, M02, M04, M06, M1, and M3.
The circle and crosses represent the Hill and Jacobi radii, respectively.
}
\label{fig:10}
\end{center}
\end{figure}

The distributions of the average specific angular momentum $j_r$ for different models are shown in Figure~\ref{fig:10}.
The circles in this figure mean the Hill radii $\rh$.
The crosses indicate the Jacobi radii $r_{\rm J}$ inside which gas is considered to be bound by the gravity of the protoplanet.
We determine the Jacobi radii from the contour of the Jacobi energies as in Figures~\ref{fig:7} and \ref{fig:8} lower panels.
The Jacobi radii in all models are distributed in the range of $r_{\rm J} \simeq 0.5-1 \rh$.
Figure~\ref{fig:10} indicates that a more massive protoplanet has an envelope with larger amount of the specific angular momentum.
Thus, the specific angular momentum accreting onto the protoplanet system increases as the protoplanet mass increases.
The rapid drops at large radii indicate the reverse of the rotation axis as shown in Figures~\ref{fig:3} and \ref{fig:9}.
In Figure~\ref{fig:10}, in model M01 ($M=0.1\mj$), $j_r$ at $r=\rh$ is twice of that at $r=r_{\rm J}$, while, in models M02, M04, M06, M1, and M3 ($M> 0.2\mj$), there are little differences between the average specific angular momentum derived from the Hill radius $j_{\rh}$ and that derived from Jacobi radius $j_{r_{\rm J}}$.
Thus, we can safely estimate the average specific angular momentum using either the Hill radius or Jacobi radius for models with $M>0.2\mj$.

\subsection{Evolution of the Specific Angular Momentum}
\label{sec:dis-evo}
To properly calculate the angular momentum of the protoplanet system,  we have to calculate the planetary growth from the solid core with $\simeq 10\me$ to the present mass.
However, it takes huge computation time to calculate all evolution phases.
Thus, we estimate the angular momentum of the protoplanet system according to the following procedure:
(i) we calculate the average specific angular momenta of the gas flowing into the protoplanet system at the fixed masses of the protoplanet (i.e., under the same parameter $\rht$) in models changing the mass of the protoplanet, then
(ii) derive the relation between the average specific angular momentum and mass of the protoplanet, and describe it as a function of the protoplanet mass, and  
(iii)  estimate the angular momentum of the protoplanet system integrating the average specific angular momentum by mass up to the present value of gas giant planets.

At first, we analytically estimate the specific angular momentum of the protoplanet system, then compare it with numerical results.
We assume that the gas that overcomes the Hill potential flows into the Hill sphere (or a protoplanet system) with the Kepler velocity of the protoplanet at the Hill radius $\rh$. 
When the protoplanet mass is $M_{\rm p}$, the Kepler velocity at $\rh$ is given by 
\begin{equation}
v_{{\rm K}, \rh} = \sqrt{\dfrac{G M_{\rm p}}{\rh} }  = \sqrt{3} \Omega_{\rm p} \rh \propto M^{1/3}.
\label{eq:kepler}
\end{equation}
Note that $\Omega_{\rm p}$ is constant when $\ro$ is fixed.
The specific angular momentum can be written as  
\begin{equation}
j = \sqrt{3} \Omega_{\rm p} \rh^2.
\end{equation}
Thus, the specific angular momentum is proportional to $ \propto \rh^2$.
In addition, when the mass of the central star is fixed, the specific angular momentum is proportional to 
\begin{equation}
j \propto M_{\rm p}^{2/3}.
\label{eq:ana-j}
\end{equation}
Thus, the specific angular momentum increases with 2/3 power of the protoplanet mass.

\begin{figure}
\begin{center}
\includegraphics[width=150mm]{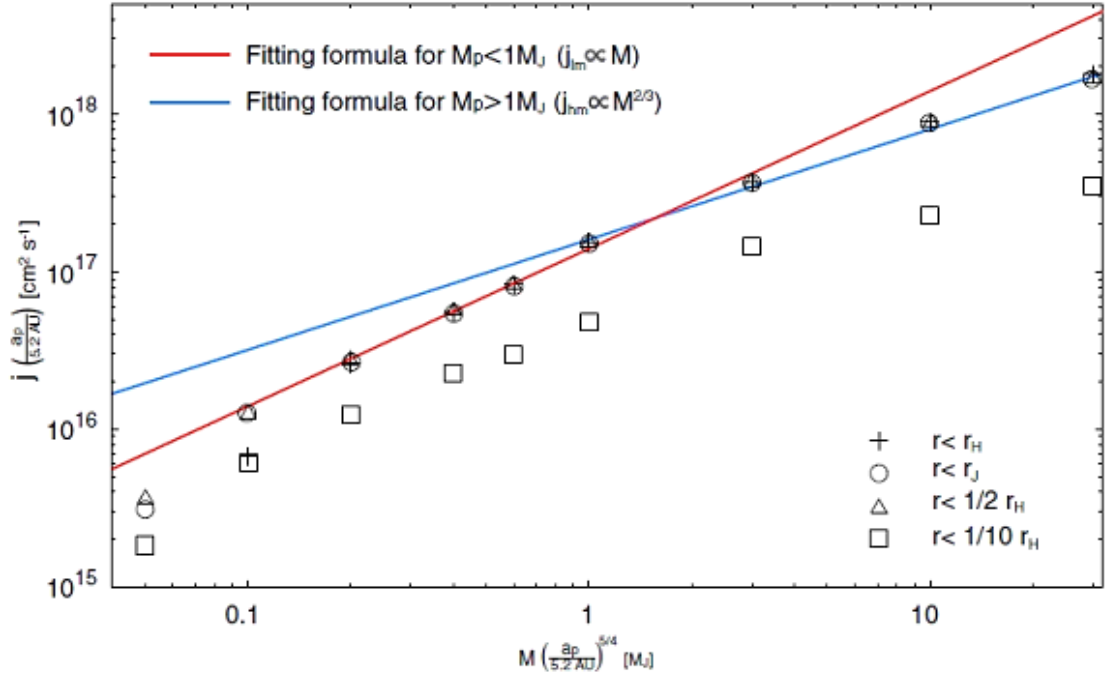}
\caption{
Specific angular momenta in the region of  $r < \rh$ ($+$), $r<r_{\rm J}$ ($\circ$), $r<1/2\,\rh$ ($\triangle$), and $r<1/10\,\rh$ ($\square$) against the protoplanet mass.
Blue and red lines are the fitting formulae for $M>1\mj$ (blue) and $M<1\mj$ (red), respectively.
}
\label{fig:11}
\end{center}
\end{figure}

Figure~\ref{fig:11} shows the average specific angular momentum derived from all models against the protoplanet mass.
The average specific angular momenta are estimated in the region of $r < \rh$ ($+$) , $r < 0.5\rh$ ($\triangle$), $r < 0.1\,\rh$ ($\square$), and $r < r_{\rm J}$ ($\circ$). 
Although we fixed the protoplanet mass in each model, we can consider the horizontal axis in Figure~\ref{fig:11} as time sequence of the protoplanet system.
Figure~\ref{fig:11} clearly shows that more massive protoplanet can acquire the envelope with larger average specific angular momentum.
Although the average specific angular momenta for $r<0.5\,\rh$ differ from those for $r<\rh$ for models with $M_{\rm p}<0.2\mj$, there are little difference for models with $M_{\rm p}>0.2\mj$.
In addition, when we adopt the average specific angular momenta in the region of $r<0.1\,\rh$, we underestimate them for any model as shown in  Figure~\ref{fig:11}.
Therefore, we can properly estimate the average specific angular momentum of the protoplanet system in any region of $r < 0.5 - 1 \rh $ for models with $M_{\rm p}>0.2M_{\rm J}$.

In Figure~\ref{fig:11}, the red and blue lines are fitting formulae of the evolution of the average specific angular momentum as a function of the protoplanet mass for $M_{\rm p}<\mj$ (low mass, $j_{\rm lm}$; red line), and $M_{\rm p}>\mj$ (high mass, $j_{\rm hm}$: blue line).
They are given by 
\begin{equation}
j_{\rm lm} = 1.4\times10^{17} \left(\dfrac{M_{\rm p}}{1\mj} \right) \ \  \jcm \hspace{3mm} \mbox{for $M_{\rm p} < 1\mj$},  \\
\label{eq:jlm}
\end{equation}
and
\begin{equation}
j_{\rm hm} = 1.6\times10^{17} \left(\dfrac{M_{\rm p}}{1\mj} \right)^{2/3} \jcm \hspace{3mm} \mbox{for $M_{\rm p} > 1 \mj$}.
\label{eq:jhm}
\end{equation}
Figure~\ref{fig:11} shows that the evolution of the average specific angular momentum for $M_{\rm p}\gtrsim1\mj$ is well described by $j \propto M_{\rm p}^{2/3}$  which corresponds to the equation~(\ref{eq:ana-j}).
On the other hand, the growth rate of the average  specific angular momentum for $M_{\rm p}<1\mj$ is larger than $j \propto M_{\rm p}^{2/3}$, and can be fitted by $j\propto M_{\rm p}$.
We ignored the thermal effect  when we derive equation~(\ref{eq:ana-j}).
When $M_{\rm p}$ is small, the gas flowing into the protoplanet system is affected relatively strongly by the thermal pressure.
We can estimate the mass at which the gravity dominates the thermal pressure force from the balance between the thermal pressure gradient and gravitational forces.
Near the Hill radius, the thermal pressure gradient force is more dominant than the gravity when the Kepler speed is slower than the sound speed (i.e., $v_{\rm K} < c_{\rm s}$).
On the other hand, when $v_{\rm K} > c_{\rm s}$, the gas flow is controlled mainly by the gravity of the protoplanet even near the Hill radius.
Using equation~(\ref{eq:kepler}), $v_{\rm K} = c_{\rm s}$ is realized when $\rh = c_{\rm s} /(\sqrt{3}\Omega_{\rm p})$ which corresponds to $M = 0.08\mj$ of the protoplanet mass at $\ro = 5.2$\,AU.
Thus, the gas flow is largely affected by the thermal pressure for $M_{\rm p} \ll 0.08\mj$, while it is not so affected by the thermal pressure for $M_{\rm p} \gg 0.08\mj$.
In Figure~\ref{fig:11}, the evolution of the angular momentum for $M_{\rm p}>1\mj$ well corresponds to the analytical solution ($j\propto M_{\rm p}^{2/3}$) indicating  that the thermal effect is negligible for $M_{\rm p} \gtrsim 1\mj$.
However, for $M_{\rm p} < 1\mj$, the average specific angular momentum is smaller than $j_{\rm hm}$ (blue line).
Thus,  when the protoplanet is $M_{\rm p}<1\mj$, the thermal effect is not negligible for the acquisition process of the angular momentum because the gravitational potential is relatively shallow.
Thus, the thermal pressure seems to remain important for $M\lesssim 1\mj$, although $v_{\rm K} = c_{\rm s}$ realized at $M=0.08\mj$.

\begin{figure}
\begin{center}
\includegraphics[width=150mm]{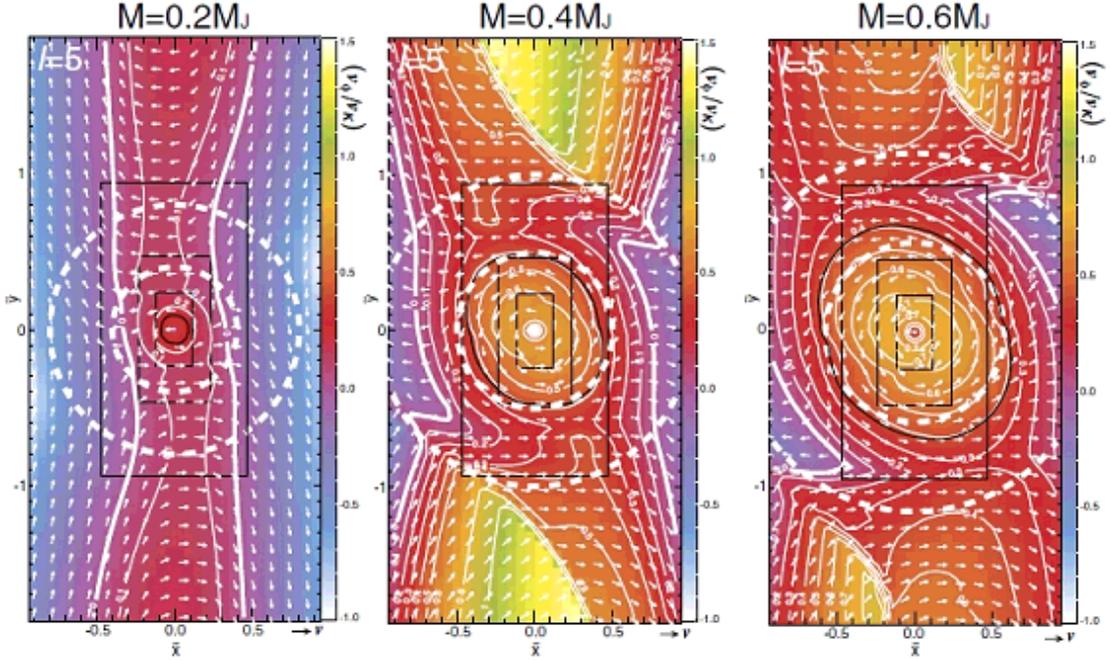}
\caption{
The ratio of the azimuthal to the Kepler velocity  $v_{\phi}/v_{\rm K}$ ({\it color} and {\it contours}) around the protoplanet for models M02 (left), M04 (middle), and M06 (right).
Three levels of grids are superimposed in each panel.
The outer and inner white-dashed-circles represent the Hill radius $\tl{r}=\rht$ (outer) and the half of the Hill radius $\tl{r}=1/2\rht$ (inner), respectively.
The black circle represents the contour of $v_{\phi}=c_{\rm s}$, inside which gas rotates with supersonic velocity $v_{\rm \phi}>c_{\rm s}$.
}
\label{fig:12}
\end{center}
\end{figure}

To verify the relation of the thermal and gravitational effects, in Figure~\ref{fig:12}, we plot the ratio of the azimuthal to Kepler velocity ($v_{\phi}/v_{\rm K}$) around the protoplanet on the midplane for models M02, M04, and M06.
In this figure, closed contours inside the Hill radius indicate that gas revolves around the protoplanet.
For example, the closed contour of $v_{\rm \phi} / v_{\rm K} =0.5$ in Figure~\ref{fig:12} means that gas rotates along the contour with 50\% of the Kepler velocity.
The black circles represent contours of $v_{\rm \phi} = c_{\rm s}$, inside which gas revolves around the protoplanet with supersonic velocity. 
Thus, the gravitational force of the protoplanet is dominant inside the black circles, while the thermal pressure gradient force is dominant outside the black circles.
Figure~\ref{fig:12} shows that the radius of the $v_{\rm \phi} = c_{\rm s}$ contour increases with the protoplanet mass indicating that the region that dominated by gravity of the protoplanet extends with the protoplanet mass.
As the protoplanet mass increases, the flow speed inside the Hill radius approaches to the Kepler velocity, and the region dominated by the gravity of the protoplanet spreads outward.
In this way, since the thermal effect decreases as the protoplanet mass increases, the growth rate of the average specific angular momentum approaches to the analytical solution (eq.~\ref{eq:ana-j}).

\section{DISCUSSION}
\subsection{Effect of Thermal  Pressure}
\label{sec:thermal}
In this paper, we investigated the evolution of the protoplanet system under the isothermal approximation, which might not be problematic in the circumplanetary disk, but may not be valid in the very proximity to the protoplanet.
To investigate the thermal effect around the protoplanet, we adopted the sink cell in some models.
As shown in \S\ref{sec:sink}, the cell width in the fiducial model having $l_{\rm max}=8$ is about 4 Jovian radius.
Thus,  gas inside $r < 4 \rj$  has the same thermal energy.
If the protoplanet has almost the same size as the present gas giant planet, the thermal energy around the protoplanet may be overestimated in model without the sink cell, because an actual protoplanet is embedded in the small part of the innermost cell.
On the other hand, when we adopt the sink cell, the thermal energy around the protoplanet is underestimated, because the thermal energy is artificially removed from the sink cell.
However, as shown in \S\ref{sec:sink}, when the sink radius is much smaller than the Hill radius ($r_{\rm sink} \ll 1/100 \rh$), there are little differences in the angular momentum acquired by the protoplanet system between models with and without the sink cell.
This is because a large part of the angular momentum of the protoplanet system is distributed around the Hill radius ($0.5\,\rh\lesssim r \lesssim \rh$) as shown in \S\ref{sec:res-ang}.
The Jovian radius is much smaller than the Hill radius ($\rh = 744\, \rj$).
In addition, the angular momentum flowing into the protoplanet system is determined by the shearing motion in the region of  $r \simeq \rh$ (see \ref{sec:dis-evo}).
Therefore, under the assumption that the protoplanet is smaller than the innermost cell width, it is expected that the thermal effect from the protoplanet is sufficiently small for the acquisition process of the angular momentum when we are using  sufficient smaller cells than the Hill radius.

\citet{mizuno80}, \citet{bodenheimer86}, and \citet{ikoma00} suggested that gaseous protoplanet has a large envelope with high temperature.
For example, \citet{mizuno80} showed that, for Jovian case, gas distributed in the range of $r\gtrsim 2\times 10^{11}$\,cm behaves isothermally, while that distributed in the range of $r \lesssim 2\times 10^{11}$\,cm behaves adiabatically in their spherically symmetric calculations (see Fig.4 of \citealt{mizuno80}).
Thus,  gas in the range of $r \lesssim 2\times 10^{11}$\,cm ($\sim30\,\rj$) has higher temperature than the ambient medium.
However, it is expected that the gaseous envelope hardly affect the angular momentum of the protoplanet system, because the size of the envelope ($\sim30\,\rj$) is much smaller than the Hill radius ($\rh = 744\, \rj$).
A large part of the angular momentum is distributed in the range of $0.5\rh \lesssim r \lesssim \rh$, as shown in \S\ref{sec:res-ang}.
On the other hand, the size of the circumplanetary disk may be affected by the thermal envelope.
Thus, when we investigate the formation of the circumplanetary disk, we have to include the realistic thermal evolution around the protoplanet.
However, it is difficult to study the thermal evolution around the protoplanet, because we need to solve the radiation hydrodynamics in three dimensions. 
In a subsequent paper, we will investigate the effect of the thermal envelope under simple assumptions.

\subsection{Angular Momentum for Jupiter and Saturn}
\label{sec:jupiter-saturn}
In \S\ref{sec:dis-evo},  we fixed the physical quantities as those at $\ro=5.2$\,AU.
However,  in our calculation, since we use the dimensionless quantities, we can rescale those at any  orbits $\ro$.
Under the standard model \citep{hayashi85}, we can generalize equation~(\ref{eq:jlm}) as
\begin{equation}
j_{\rm lm} = 7.8 \times 10^{15} \left(\dfrac{M_{\rm p}}{M_{\rm J}} \right) \left(\dfrac{\ro}{1\,{\rm AU}}  \right)^{7/4} \jcm 
  = 2.5 \times 10^{13} \left( \dfrac{M_{\rm p}}{\me} \right)   \left( \dfrac{\ro}{1\, {\rm AU}} \right)^{7/4} \jcm.
\label{eq:fit3} 
\end{equation}
We assume that the protoplanet system acquires the gas with the average specific angular momentum of equation~(\ref{eq:fit3}) when the protoplanet has a mass of $M_{\rm p}<1\mj$.
Thus, to estimate the angular momentum of the protoplanet system, we need to integrate equation~(\ref{eq:fit3}) by the mass until the present values of gas giant planets.
For example, the angular momentum in our model for Jupiter is 
\begin{equation}
J_{\rm J} = \int^{\mj} j_{\rm lm} (5.2\,{\rm AU})\, dM = 1.3 \times 10^{47} {\rm g\jcm},
\end{equation}
which is about 30 times larger than that of present Jupiter ($4.14\times 10^{45}$g$\jcm$).
On the other hand, the angular momentum in our model for Saturn is 
\begin{equation}
J_{\rm S} = \int^{M_{\rm s}} j_{\rm lm} (9.6\,{\rm AU}) \, dM = 3.6\times 10^{46} \ \ {\rm g\,\jcm},
\end{equation}
which is 50 times larger than that of present Saturn ($7.2 \times 10^{44}$g$\jcm$), where $M_{\rm s} = 95.16\me$ is the Saturnian mass.
Note that the orbital angular momenta of the {\em present} Jovian and Saturnian satellites can be ignored in the protoplanet system because they are considerably smaller  than the Jovian and Saturnian spin angular momenta.

The above estimation corresponds to the angular momentum of proto planet-disk system, which includes the central planet and protoplanetary disk.
We showed that the angular momentum of the protoplanet system is $30-50$ times larger than  present spin of the gaseous planet.
The angular momenta flowing into the Hill sphere are distributed into the spin of the protoplanet and orbital motion of the circumplanetary disk.
It is expected that a large fraction of the total angular momentum contributes to the formation of the circumplanetary disk and the residual contributes to the spin of the planet.
Although the fraction of the angular momentum of the circumplanetary disk to the spin of protoplanet is not correctly estimated in our calculation, a part of the angular momentum is certainly distributed into the disk.
Thus, the angular momentum transfer and dissipation mechanism for the circumplanetary disk are necessary to follow further evolution of the protoplanet system.
\citet{takata96} proposed the despin mechanism of the protoplanet by planetary dipole magnetic field, in which an initially rapidly rotating protoplanet can be spun down to the present value by the magnetic interaction between the protoplanet and circumplanetary disk.

\subsection{Disk formation and Implication for Satellite Formation}
\label{sec:dis-disk}
When the circumplanetary disk is formed around the protoplanet, it is possible to form satellites in the disk.
While there are many scenarios for the satellite formation \citep[e.g.,][]{stevenson86},  regular satellites around gas giant planets are supposed to be formed in the gaseous disk as for the planet formation in the protoplanetary disk \citep[e.g.,][]{korycansky91, canup02}.
Observations showed that the regular satellites around Jupiter and Saturn are distributed only in the close vicinity of the planet ($r\lesssim 50\,\rp$), and are on prograde orbits near the equatorial plane.
Thus, it is expected that these regular satellites  formed in the circumplanetary disk.

We discussed the angular momentum of the protoplanet system in \S\ref{sec:dis-evo}.
As the angular momentum flowing into the Hill sphere is brought into both the protoplanet and circumplanetary disk, we cannot estimate the fraction of angular momentum for the circumplanetary  disk. 
When we assume that the centrifugal force is balanced with the gravity of the protoplanet, we can derive the centrifugal radius $r_{\rm cf}$ as 
\begin{equation}
r_{\rm cf} = \dfrac{j^2}{GM_{\rm p}}.
\label{eq:cent}
\end{equation}
To quantify the disk size, we adopt the specific angular momentum in equation~(\ref{eq:cent}) as those in equation~(\ref{eq:fit3}).
The centrifugal radius for the proto-Jovian disk with $M_{\rm p} = 1 \mj$ and $\ro =5.2$\,AU is
\begin{equation}
r_{\rm cf, J}  = 1.5 \times10^{11} \ \ {\rm cm},
\label{eq:r-est}
\end{equation}
which is 22 times as large as Jovian radius ($r_{\rm Jup} = 7.1\times 10^9\cm$).
The Galilean satellites, which are regular  satellites, are distributed in the range of $6\,r_{\rm Jup}\lesssim r \lesssim 27\,r_{\rm Jup}$, which is consistent with the centrifugal radius derived from our calculation.
Note that the disk size is expected to be larger than equation~(\ref{eq:r-est}) owing to the thermal effect around the protoplanet. 
In the same way, we estimate the centrifugal radius of proto-Saturnian disk as
\begin{equation}
r_{\rm cf, S}  = 4.1\times10^{11} \ \ {\rm cm},
\end{equation}
which is 68 times as large as Saturnian radius ($r_{\rm Sat}=6.0\times10^9$\,cm).
The Saturnian representative regular satellites (Mimas, Enceladus, Tethys, Dione, Rhea, Titan, Hyperion, and Iapetus) are distributed in the range of $3\, r_{\rm Sat} \lesssim  r \lesssim 60\, r_{\rm Sat}$, which is also corresponding to the centrifugal radius of our result.

\begin{figure}
\begin{center}
\includegraphics[width=150mm]{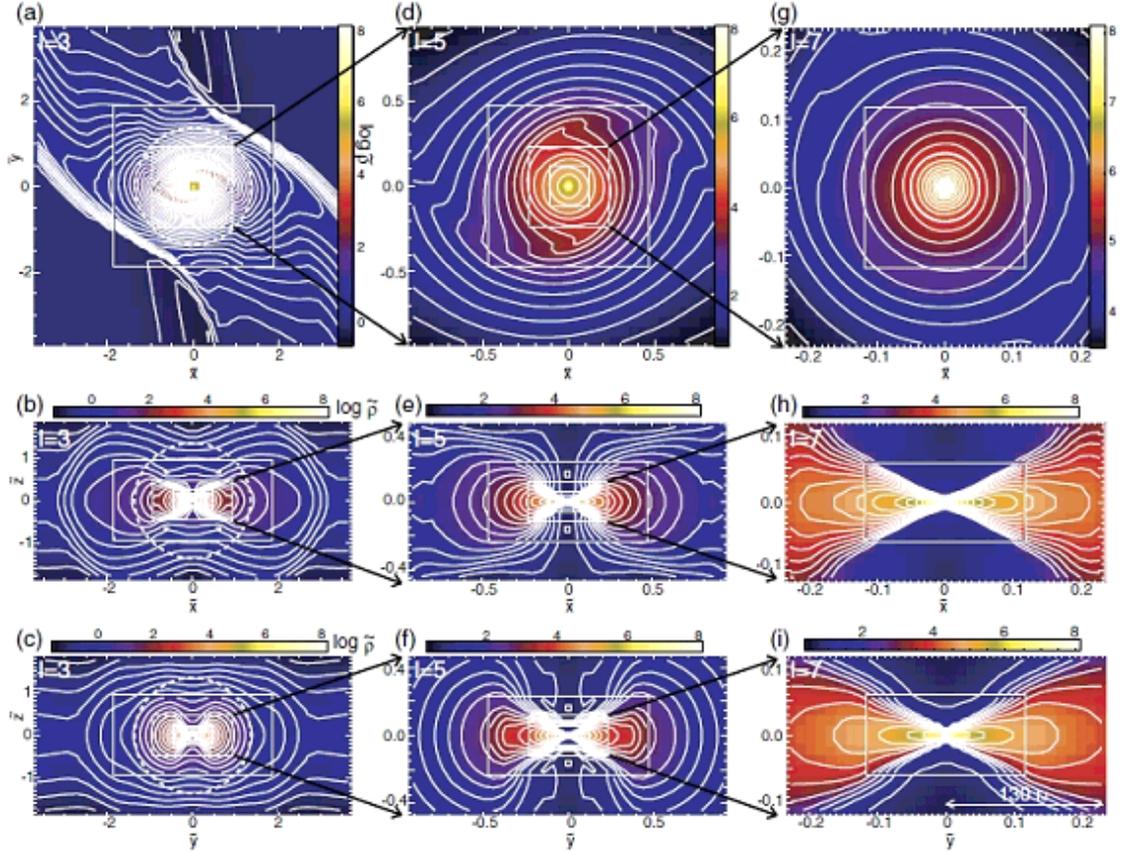}
\caption{
Density distribution ({\it color} and {\it contours}) on the cross section in the $\tl{z}=0$ (upper panels), $\tl{y}=0$ (middle panels), and $\tl{x}=0$ (lower panels) plane with different outermost grid levels $l=3$ (left), 5 (right) and 7 (right) for model M1.
Three grid levels are superimposed in each panel.
Right panels {\it g}, {\it h}, and {\it i} are eight times enlargement of left panels {\it a}, {\it b}, and {\it c}.
The dotted circles represent the Hill radii $\tl{r}=\rht$.
}
\label{fig:13}
\end{center}
\end{figure}

Figure~\ref{fig:13} shows the structure around the protoplanet at the end of the calculation ($\deft \simeq 20$) for model M1, in which the protoplanet has mass of $1\mj$.
Each right panel is 4 times magnification of each left panel.
Figures~\ref{fig:13}{\it a}, {\it b}, and {\it c} show the shock fronts outside the Hill radius, and a thick disk between the shock front and the Hill sphere.
In the proximity to the protoplanet, a concave structure is seen in  Figures~\ref{fig:13}{\it e} and {\it f}.
The contours in these figures rapidly drop around the rotation axis (i.e., $z$-axis). 
Thus, even inside the Hill sphere, except for the central region, the disk has a thick torus-like structure.
Figure~\ref{fig:13}{\it g}, {\it h}, and {\it i} show the structure in the very proximity to the protoplanet.
On the midplane, the elliptical structure is seen in a large scale (Fig.~\ref{fig:13}{\it d}), while an almost round shape is seen in the proximity to the protoplanet (Fig.~\ref{fig:13}{\it g}) because the force acting on gas in this region is dominated by the gravity of planet.
The contours in Figures~\ref{fig:13}{\it h} and {\it i} show  a very thin disk in the region of $\tl{r} \lesssim 0.1$ (or $r \lesssim 50\,\rj$).
Thus, the regular satellites might be formed in these circumplanetary disks.
However, to study the satellite formation in more detail, we need more realistic calculations.

\section{SUMMARY}
To study the gas flow pattern and angular momentum accretion onto the
 protoplanet system, we have calculated the evolution of the protoplanet
 system in the circumstellar disk.
Firstly, we have investigated the dependence of the angular momentum
 accreting onto the protoplanet system on spatial resolution with
 different cell widths, and confirmed its convergence for the cell width
 much smaller than the Hill radius.  
Next, adopting the sink cell whose size is comparable to or slightly
 larger than the radius of the present gaseous planets in the solar
 system, we have checked that, in both models with and without the sink, the
 thermal effect around the protoplanet barely affects the gas flow
 pattern and angular momentum of the protoplanet system. 
Thirdly, with sufficiently high spatial resolution, we have calculated
 the evolution of the protoplanet system with the protoplanet mass in
 the range of $0.05\mj\le M_{\rm p} \le30\mj$, where $M_{\rm  p}$ and
 $\mj$ are the protoplanet and Jovian masses, respectively.  
The following results are obtained:
\begin{itemize}
\item The gas flow pattern in three dimensions is qualitatively
      different from that in two dimensions: the gas is flowing onto the
      protoplanet system mainly in the vertical direction in
      three-dimensional simulations. 
\item  The specific angular momentum increases as $j\propto M_{\rm p}$
      when the protoplanet mass is  $M_{\rm p}\lesssim 1\mj$,  while it
      increases as $j\propto M_{\rm p}^{2/3}$ when $M_{\rm p}\gtrsim
      1\mj$. 
\item The angular momentum of the protoplanet system is $30-50$ times
      larger than the present spin of the gaseous planets in the solar
      system.  
\item A thin disk is formed only in the region of $r\lesssim 20-60\,\rp$, where $\rp$
      is the radius of the planets.
      This location agree with the orbital radii of regular satellites
      around the present gaseous planets in the solar system. 
\end{itemize}
These conclusions imply that the protoplanet system can acquire
 sufficient angular momentum from the circumstellar disk, and despin
 mechanisms of the protoplanet system are necessary to realize the
 present spin of the gaseous planets in the solar system.
We are planning higher-resolution simulations to further investigate the
 evolution of the angular momentum of the gas giant planets in  near
 future.

\acknowledgments

We have greatly benefited from discussion with H. Tanaka, and T. Muto. 
We also thank T. Hanawa for contribution to the nested-grid code.
Numerical calculations were carried out with Fujitsu VPP5000 at the
 Center for Computational Astrophysics, the National Astronomical
 Observatory of Japan. 
SI is grateful for the hospitality of KITP and interactions with the participants of the program ``Star Formation through Cosmic Time''.
This work is supported by the Grants-in-Aid from MEXT (15740118, 16077202, 16740115, 18740104). 


\begin{table}   
\caption{Model parameters}
\label{table:table1}
\begin{center}
\begin{tabular}{cccccccccccccc}
\hline
Model & $\rht$   &  $M_{\rm p}$  {\footnotesize (5.2 AU)} 
\tablenotemark{a}
& $l_{\rm max}$ & $\Delta \tl{s}$ ($10^{-3}$)  & $\tl{r}_{\rm sink}$ ($10^{-2}$)

\\
\hline
M005  & 0.5   & 0.05  & 8 &$7.3$ & --- \\
M01   & 0.63  & 0.1   & 8 &$7.3$ & --- \\
M02   & 0.8   & 0.2   & 8 &$7.3$ & --- \\
M04   & 1.0   & 0.4   & 8 &$7.3$ & --- \\
M06   & 1.15  & 0.6   & 8 &$7.3$ & --- \\
M1    & 1.36  & 1     & 8 &$7.3$ & --- \\
M3    & 1.95  & 3  & 8 &$7.3$ & --- \\
M10   & 2.91  & 10  & 8 &$7.3$ & --- \\
M30   & 4.21  & 30  & 8 &$7.3$ & --- \\
M04L5 & 1.0   & 0.4   & 5 &$5.9$ & --- \\
M04L6 & 1.0   & 0.4   & 6 &$2.9$ & --- \\
M04L7 & 1.0   & 0.4   & 7 &$1.5$ & --- \\
M04L9 & 1.0   & 0.4   & 9 &$3.7$ & --- \\
M04L10& 1.0   & 0.4   & 10&$1.8$ & --- \\
M04S01& 1.0   & 0.4   & 8 &$7.3$ & $1$ \\
M04S03& 1.0   & 0.4   & 8 &$7.3$ & $3$ \\
\hline
\end{tabular}
\end{center}
\tablenotetext{a}{in unit of Jupiter mass  $M_{\rm J}$}
\end{table}

\end{document}